\documentclass[a4paper,11pt]{article}
\pdfoutput=1 %
\usepackage{jcappub}
\usepackage{rotating}
\usepackage{array}
\usepackage{amsmath,bm}
\usepackage[normalem]{ulem}
\usepackage{slashed}
\usepackage{booktabs}
\usepackage[pdftex,table]{xcolor}
\usepackage{units}
\usepackage{subcaption}
\usepackage{comment}
\usepackage{caption,ulem}
\usepackage[T1]{fontenc} 
\usepackage{array}
\usepackage{amsmath,bm}
\usepackage{sectsty}
\allsectionsfont{\boldmath}

\usepackage{tabularx}

\newcolumntype{Y}{>{\centering\arraybackslash}X}

\DeclareUnicodeCharacter{2212}{-}

\arxivnumber{LAPTH-051/24}

\title{\boldmath Disclosing the catalog pulsars dominating the Galactic positron flux}

\author[a,b]{Luca Orusa,}
\author[c]{Silvia Manconi,}
\author[d,e]{Fiorenza Donato}
\author[e]{and Mattia Di Mauro}

\affiliation[a]{Department of Astrophysical Sciences, Princeton University, \\ Princeton, NJ 08544, USA}
\affiliation[b]{Department of Physics and Columbia Astrophysics Laboratory, Columbia University, \\ New York, NY 10027, USA}

\affiliation[c]{Laboratoire d’Annecy-le-Vieux de Physique Théorique (LAPTh),\\
CNRS, USMB, F-74940 Annecy, France}

\affiliation[d]{Dipartimento di Fisica, Università di Torino,\\
  Via P. Giuria 1, Torino, Italy}

\affiliation[e]{Istituto Nazionale di Fisica Nucleare, Sezione di Torino,\\
Via P. Giuria 1, 10125 Torino, Italy}

\emailAdd{luca.orusa@princeton.edu}
\emailAdd{manconi@lapth.cnrs.fr}
\emailAdd{donato@unito.it}
\emailAdd{mattia.dimauro@to.infn.it}

\abstract{The cosmic-ray flux of positrons is measured with high precision by the space-borne particle spectrometer AMS-02. The hypothesis that  pulsars and their nebulae can significantly contribute to the excess of the AMS-02 positron flux has been consolidated after the observation of a $\gamma$-ray emission at GeV and TeV energies of a few degree size around a few sources, that provide indirect evidence that electron and positron pairs are accelerated to very high energies from these sources. 
By modeling the emission from pulsars in the ATNF catalog, we find that combinations of positron emission from cataloged pulsars and secondary production can fit the observed AMS-02 data. Our results show that a small number of nearby, middle-aged pulsars, particularly B1055-52, Geminga (J0633+1746), and Monogem (B0656+14), dominate the positron emission, contributing up to 80\% of the flux at energies above 100 GeV. From the fit to the data, we obtain a list of the most important sources for which we recommend multi-wavelength follow-up observations, particularly in the $\gamma$-ray and X-ray bands, to further constrain the injection and diffusion properties of positrons.
}

\begin{document}
\maketitle
\flushbottom
\section{Introduction}

The AMS-02 experiment has measured the cosmic positron ($e^+$) flux with unprecedented precision over the energy range of 0.5 to 1000 GeV \cite{PhysRevLett.122.041102}, although the theoretical interpretation of the flux remains a topic of debate.

At low energies, below 10 GeV, the flux is likely dominated by secondary production, which occurs through interactions between cosmic rays (CRs) and the atoms in the interstellar medium (ISM) \cite{Delahaye:2008ua,Boudaud:2014dta,DiMauro:2014iia,Evoli:2020szd,Di_Mauro_2023}. Current models suggest that secondary $e^+$ production accounts almost entirely for the flux observed by AMS-02 between 0.5 and 1 GeV and can explain between 50-70\% of the measurements at 10 GeV, depending on the assumed vertical size of the diffusive halo \cite{Di_Mauro_2023}. 
Above 10 GeV, the presence of a $e^+$ excess becomes even more pronounced, and between 100 and 1000 GeV, secondary production accounts for only around 10\% of the observed data.
The difference between the secondary flux and the observed data cannot be explained by uncertainties in propagation effects or in the $e^+$ production cross sections. In particular, the latter are estimated to bring only $5-7$\% to the overall uncertainty \cite{Orusa_2022}. Therefore, it is likely that primary sources of $e^+$ already contribute to the flux below 10 GeV. 

Pulsars are considered one of the most promising Galactic sources of high-energy $e^+$. These rotating neutron stars are believed to produce electron and positron pairs ($e^\pm$) through the spin-down mechanism, accelerating $e^+$ to energies that can reach up to PeV \cite{Bykov:2017xpo,Amato:2020zfv}. Recently, $\gamma$-ray halos at TeV energies, extending a few degrees around middle-aged pulsars, have been detected by Milagro \cite{2009ApJ...700L.127A}, HAWC \cite{Abeysekara:2017science,2023ApJ...944L..29A}, HESS \cite{HESS:2023sbf}, and LHAASO \cite{2021PhRvL.126x1103A,cao2024}. A very extended halo around Geminga has also been observed at GeV energies using Fermi-LAT data \cite{DiMauro:2019yvh}. Additionally, studies of the HESS Galactic plane survey and the latest HAWC catalogs suggest that such halos may be a common feature among Galactic pulsars \cite{Linden:2017vvb,DiMauro:2019hwn,DiMauro:2020jbz,Hooper:2021kyp}. Efforts to detect extended emissions around the Geminga pulsar using X-ray data from XMM, Chandra, NuSTAR, and SRG/\textit{eRosita} have so far been inconclusive \cite{Liu:2019sfl,Khokhriakova:2023rqh,Manconi:2024wlq}. These analyses have, however, placed upper limits on the magnetic field around Geminga, suggesting it is likely below $\mu$G strength.

The observations of $\gamma$-ray halos around pulsars provide indirect evidence that $e^\pm$ are accelerated to very high energies from these sources. In fact, very-high-energy photons are thought to be produced by CR $e^\pm$ that escape from the pulsar wind nebula (PWN) system. After exiting the PWN and entering the ISM, the $e^\pm$ interact with low-energy photons in the interstellar radiation field (ISRF), generating high-energy photons through inverse Compton scattering (ICS) \cite{Gaensler_2003,2017hsn..book.2159S}. Despite this, the exact mechanism governing the transport of $e^\pm$ in pulsar halos, and consequently the production of $\gamma$-ray emissions, remains an open question \cite{Evoli:2018aza,Lopez-Coto:2017pbk,Liu:2019zyj,Fang:2019iym,Recchia:2021kty}. One of the main challenging aspects is that the size of $\gamma$-ray halos implies that the diffusion around pulsars should be a few orders of magnitude smaller than the average of the Galaxy in order to confine $e^{\pm}$ more efficiently within a few tens of parsecs around the object. So far, there is no consensus and no satisfactory model that can explain the presence of inhibited diffusion around pulsars \cite{Recchia:2021kty,PhysRevD.105.123008}. 
For a recent review on $\gamma-$ ray halos around pulsar we refer to \cite{Amato:2024dss}.

Theoretical models that incorporate contributions from pulsars and their nebulae (pulsar in short, in what follows) have been successful in explaining the AMS-02 $e^+$ data. These models take into account both a small number of nearby pulsars and the cumulative emission from pulsars listed in existing catalogs \cite{Mlyshev:2009twa,Boudaud:2014dta,2014JCAP...04..006D,Manconi:2016byt,Fornieri_2020,DiMauro:2020cbn}, as well as simulations of the Galactic pulsar population \cite{Cholis_2018,Evoli_2021,Manconi:2020ipm,Orusa_2021}. Several published papers agree that the cumulative distribution of Galactic pulsars can explain the entirety of the $e^+$ flux above 10 GeV if they convert between $1-10\%$ of their spin-down power into $e^\pm$ pairs.

The Australia Telescope National Facility Pulsar Catalog (ATNF) \cite{Manchester:2004bp} contains thousands of Galactic pulsars, which can account for the $e^+$ flux detected on Earth. However, $e^{\pm}$ traveling in the Galaxy are affected by propagation processes, whose main effects are energy losses and diffusion. In particular, for $e^+$ above 10 GeV, energy losses prevent distant sources from contributing significantly to the observed flux. Therefore, local sources—those within 1 kpc—are expected to account for the majority of the $e^+$ flux. Thus, local sources are the ones more likely to contribute significantly to the AMS-02 data.

In our previous paper \cite{Orusa_2021}, we conducted realistic simulations of the Galactic population of pulsars, calibrating their injection properties to fit AMS-02 $e^+$ data. Our main result was that to explain the smoothness of the data, a few nearby and middle-aged pulsars should dominate the observed flux. In this paper, we take a step forward by using the latest version of the ATNF catalog \cite{Manchester:2004bp} to identify the most promising pulsars that can contribute to the AMS-02 $e^+$ flux.

We find the list of sources by considering different assumptions for the injection spectrum of $e^{\pm}$, spin-down, and particle transport models. The list of the brightest pulsars we identify could be used to determine the injection spectrum properties through multimessenger observations in X-ray and $\gamma$ rays, from GeV up to TeV energies.

The paper is organized as follows: in Section \ref{sec:positrons} we report the details of the model, Section \ref{sec:results} we list our results, in Section \ref{sec:multi-w} we show the available multi-wavelength observations of the most interesting pulsars and, finally, in Section \ref{sec:conclusions} we draw our conclusions.

\section{Positrons from catalog Galactic pulsars}
\label{sec:positrons}

In order to predict the contribution of a given Galactic pulsar to the $e^+$ flux at Earth, a number of quantities has to been defined, which we can separate into three categories, as summarized below and detailed in the following subsections.

The first set of quantities defines the characteristics of the pulsar: the distance, age and spin-down power. These are found in pulsar catalogs, and thus for observed pulsar they are fixed, while for mock pulsar populations they must be simulated \cite{Orusa_2021}. 
The second category concerns the quantities connected to the particle emission of each pulsar, which are phenomenological parameters in our current modeling of the process. They are not directly measurable for each individual source, and their value is usually  informed by the current theoretical knowledge of particle production, acceleration and emission from pulsars as well as from multi-wavelength data on specific, representative sources. Since emission parameters are not measured for each source, we simulate them following the current state-of-the-art phenomenological knowledge of these phenomena. 
Finally, the $e^+$ released from pulsars are transported in the Galaxy before reaching the Earth, and transport parameters connected to the propagation in the ISM, including diffusion and energy losses, are needed to define the modeling of this part of $e^+$ journey. 
For this category, we assume equal transport parameters for all sources, thus assuming the properties of the ISM to be on average similar for pulsars located in the Galactic disk. 

In what follows, we identify and detail our benchmark setup for the modeling of $e^+$ emission and propagation and the required parameters for each of these categories, together with three main variations, designed to explore the consequences of the uncertainty in the emission and propagation modeling on our results.  Our main setups are summarized in Section~\ref{sec:simulations}.

\subsection{ATNF catalog}\label{sec:atnf}

The sample of pulsars considered in this work is taken from the ATNF pulsar catalog \cite{Manchester:2004bp}, version 2.1.1 \footnote{http://www.atnf.csiro.au/research/pulsar/psrcat}, which lists the observed rotation-powered pulsars, including those detected only at high energies. 
Within this version of the ATNF catalog, the distance to each pulsar is estimated using the most updated electron-density model \cite{Yao_2017}. However, the choice of electron-density model introduces unavoidably biases in the distance estimates, which can significantly impact the computation of the $e^+$ flux. 
The model in \cite{Yao_2017} could lead to artificial small distances of some pulsars, as for instance the distance to the source B1055-52, which can vary between 714 pc and 93 pc, depending on the Galactic free-electron density model used \cite{Cordes:2002wz,Yao_2017}.

Sources are selected by excluding all the millisecond pulsars, namely cutting off rotation periods $P_0<0.01$ s. The emission coming from these sources would require a different model with respect to what 
implemented in this work. Furthermore, it has been shown to be negligible through an analysis realized with simulations of the millisecond pulsar population \cite{Venter_2015}.
 
After these selection cuts, a sample of 2261 pulsar is retained with a measured value of distance, age and spin-down power. 
We recognize in the sample notable sources such as Geminga and Monogem (J0633+1746 and B0656+14), which have been considered to be main candidates to contribute significantly to the $e^+$ flux at hundreds of GeV. 
Among the selected sample, 181 pulsars have distances estimated to be less than 1 kpc, and 390 have ages between 50 kyr and $10^3$ kyr. The number of pulsars having spin-down power exceeding  $\dot{E}>10^{34}$ erg/s, so with energetics comparable to the one of Geminga and Monogem, is 287.

\subsection{Positron emission model}
\label{sec:injection}
Pulsars are rotating neutron stars with a strong surface magnetic field, and magnetic dipole radiation is believed to provide a good description for their observed loss of rotational energy \cite{Bykov:2017xpo,Amato:2020zfv}. 
We consider that both $e^{\pm}$ are continuously accelerated and injected in the ISM at a rate that follows the pulsar spin-down energy. This scenario is indeed required to generate the TeV photons detected by Milagro and HAWC for Geminga and Monogem pulsar halos \cite{Abeysekara:2017hyn, DiMauro:2019yvh, Yuksel:2008rf,Recchia:2021kty}. 
We remind here the main aspects for the emission model of $e^{\pm}$ from pulsars. The formalism is more extensively presented in~\cite{DiMauro:2019yvh,Orusa_2021}, to which we refer for further details.

The injection spectrum $Q(E,t)$ of $e^\pm$ in the ISM at energy $E$ and time $t$ is described as a broken power law \cite{Abeysekara:2017science,DiMauro:2019yvh}: 
\begin{equation}
Q(E, t) = L(t) 
{\rm e}^{-E/E_{\rm c}}
\times
\begin{cases}
(E / E_{\rm b})^{-\gamma_{\rm L}} & E < E_{\rm b} \\
(E / E_{\rm b})^{-\gamma_{\rm H}} & E \ge E_{\rm b},
\label{eq:spectrum}
\end{cases} 
\end{equation}
where the cut-off energy $E_{\rm c}$ is fixed at $10^3$ TeV, $E_{\rm b}$ is the break at energies of the order of hundreds of GeV, and $\gamma_{\rm L}$, $\gamma_{\rm H}$ are the slopes below and above the break energy. 
The broken power-law model is compatible with multi-wavelength observations of PWNe, although there are large uncertainties on specific values of these parameters for each source \cite{Torres:2014iua}. 
The magnetic dipole braking $L(t)$ is described by the function:
\begin{equation}\label{eq:L}
L(t)=\frac{L_0}{\left( 1+ \frac{t}{\tau_0}\right)^{\frac{n+1}{n-1}}}
\end{equation}
where $\tau_0$ is the characteristic time scale and $n$ defines the magnetic braking index.
The total energy emitted by the source only into $e^+$ is given by:
\begin{equation}\label{eq:Etot}
    E_{tot}=\eta W_0= \int_{0}^T dt \int_{E_1}^{\infty} dE E Q(E,t)
\end{equation}
through which we obtain the value of $L_0$, fixing $E_1$=0.1 GeV \cite{Sushch:2013tna, Buesching:2008hr}. The parameter $\eta$ encodes the efficiency of conversion of the spin-down energy into $e^+$ (which is half of the efficiency of conversion into $e^\pm$). Changing the value of $E_1$ affects the normalization of the total spectrum, which in turn impacts the efficiency $\eta$ of individual sources. For the injection spectrum parameters used in this work, as listed in Table \ref{tab::models}, setting $E_1$ to the $e^+$ mass can result in a maximum renormalization of the flux obtained with $E_1=0.1$ GeV by a factor of 0.65. This occurs for $\gamma_L = 2$ and $\gamma_F=2.8$, which represents the specific parameter set that maximizes the effect of lowering the $E_1$ value. Choosing, on the other hand, values of $E_1$ larger than 0.1 GeV would lead to a flux renormalization greater than 1, thereby requiring lower efficiencies. $W_0$ is the initial rotational energy of a pulsar and can be computed from the pulsar age $T$, the spin-down luminosity $\dot{E}$ and the decay time $\tau_0$:
\begin{equation}
 W_0 = \tau_0 \dot{E} \left( 1+ \frac{T}{\tau_0} \right)^{\frac{n+1}{n-1}}\,.
 \label{eq:W0PWN}
\end{equation} 
Here $\dot{E} = dE_{\rm rot}/dt$ is the rate at which the pulsar rotational kinetic energy is dissipated and is called spin-down luminosity.
The \textit{actual} age $T$ and the  \textit{observed} age $t_{\rm{obs}}$ are related by the source distance $d$ by $T = t_{\rm obs} + d/c$. 
It is possible to derive the value of $\tau_0$ from the initial rotation period of the pulsar $P_0$ and is first derivative, as:
\begin{equation}\label{eq:tau0}
    \tau_0= \frac{P_0}{(n-1)\dot{P}_0}.
\end{equation}

Assuming a small deviation from the dipole nature of the magnetic field $B$ of the pulsar, $\dot{P_0}$ may be obtained from  $P^{n-2}\dot{P} = a k (B \sin \alpha)^2\,$ \cite{Ridley_2010},
where the angle $\alpha>0$ describes the inclination of the magnetic dipole with respect to the rotation axis, and $a$ and $k$ are constants that depend on the canonical characteristics of neutron stars (see \cite{Chakraborty:2020lbu} for details).

\subsubsection{Simulated emission parameters}
While the age and spin-down power are characteristics available for pulsars identified in catalogs, it is not possible to derive the value of $\tau_0$, the spectral indexes and the conversion efficiency directly from the current measurements,  and assumptions about other quantities involved, e.g. in the calculation of $\tau_0$, are thus necessary. To complement the emission parameters not directly listed in pulsar catalog for each source, we proceed as follows. 

As for the typical decay time $\tau_0$, we  either assume  an unique value of $\tau_0=10$ kyr for all the sources, a reference usually adopted in literature (see e.g. \cite{Abeysekara:2017science,DiMauro:2019yvh}), or draw a distribution obtained from the simulation of artificial values of $P_0$, $n$, $B$, and $\alpha$ for each source. 
In the latter setup, in order to assign to each pulsar a $\tau_0$ value, we extended the functions implemented in the Python module {\tt gammapy.astro.population} \cite{Nigro_2019,CTAConsortium:2017xaq} to  sample the values of $P_0, B, n$ and $\alpha$ from the distributions provided in \cite{Chakraborty:2020lbu} (\texttt{CB20}).  
From these quantities, we extract the value of $\tau_0$ accordingly for each source. 
In this way, a distribution of $\tau_0$ starting from realistic distributions of the other quantities is obtained. If not stated differently, this represents the benchmark setup in our computations. 

Since the spectrum of accelerated particles is uncertain, and may vary significantly for each source \cite{Gaensler:2006ua, Mlyshev:2009twa}, we sample $\gamma_{\rm L}$ from uniform distributions within [$1.0-2.0$], $\gamma_{\rm H}$ within [$2.0-2.8$] and $E_b$ within [$300-600$] GeV. 
Finally, the value of $\eta$ for each source is sampled from a uniform distribution in the range [$0.01-0.1$].

Unless differently stated, we will consider all the pairs produced by sources with ages above 50 kyr, since $e^\pm$ accelerated to TeV energies in the termination shock are believed to be confined in the nebula or in the supernova remnant (SNR) until the merger of this system with the ISM, estimated to occur tens of kyr after the pulsar formation \cite{2011ASSP...21..624B}. 
We thus leave out young sources for which the $e^\pm$ pairs might still be confined in the parent remnant. However, this effective treatment does not account for possible spectral or time-dependent modifications of the released particles. 
To understand the consequences of this assumption on the interpretation of the  $e^+$ flux at Earth, we also test the hypothesis that only the $e^\pm$ produced after the pulsar escapes from the SNR contribute to the flux at the Earth. 
Following \cite{Evoli_2021,Orusa_2021}, we define $t_{BS}$ as the time at which the source leaves the parent SNR due to its proper motion and eventually forms a bow-shock nebula. The escape time of the pulsar from the remnant is described by:
\begin{equation}
    t_{BS} \simeq 56 \left(\frac{E_{SN}}{10^{51} \text{erg}}\right)^{\frac{1}{3}} \left( \frac{n_0}{3 \text{ cm}^{-3}}\right)^{-\frac{1}{3}} \left( \frac{v_k}{280 \text{ km/s}} \right)^{-\frac{5}{3}} \text{kyr}
    \label{eq:bow_shock}
\end{equation}
where $n_0$ is the ISM density around the SNR taken to be 3 cm$^{−3}$, $E_{SN}$ = $10^{51}$ erg is the energy emitted by the SN explosion and $v_k$ is the birth velocity of the pulsar. The formalism is reported in~\cite{van_der_Swaluw_2003}, to which we refer for further details. The parameter values are taken from \cite{Evoli_2021} to ensure consistency with the analysis framework used in that study.
To test the scenario described by Eq.~\eqref{eq:bow_shock}, for each source its birth-kick velocity is simulated, adopting its distribution as reported in \cite{2006ApJ...643..332F} (\texttt{FK06VB}) and implemented in  {\tt gammapy.astro.population} \cite{Nigro_2019,CTAConsortium:2017xaq}, which is the sum of two Gaussians (see their Eq.~7) for each of the three velocity components.

\subsection{Positron propagation to the Earth and anisotropy}
\label{sec:propagation}

Once charged particles are injected into the Galaxy, they can propagate and eventually reach the Earth. Here, we briefly outline our treatment of Galactic propagation of $e^\pm$ and refer the reader to ~\cite{2010A&A...524A..51D,Manconi:2018azw,Manconi:2020ipm,Orusa_2021} for further details. 
The number density per unit time and volume, $N_e(E, {\bf r}, t)$, of $e^\pm$ at  observed energy $E$, position ${\bf r}$ in the Galaxy, and time $t$, which is the solution to the propagation equation considering only an homogeneus diffusion and energy losses, is given by \cite{DiMauro:2019yvh}:
\begin{equation}\label{eq:numberdensity}
    \mathcal{N}_e(E, {\bf r}, t)=\int_0^t dt' \frac{b(E_s)}{b(E)} \frac{1}{(\pi \lambda^2(t',t,E))^{\frac{3}{2}}}  \exp \left(-\frac{| {\bf r}-{\bf r_s}|^2}{\lambda ^2 (t',t,E)} \right) Q(E_s,t')
\end{equation}
where the integration over $t'$ accounts for the source releasing $e^{\pm}$ continuously in time. The energy $E_s$ is the initial energy of $e^{\pm}$ that cool down to $E$ in a loss time $\Delta \tau$:
\begin{equation}\label{eq:losstime}
    \Delta \tau \equiv \int_E^{E_s} \frac{dE'}{b(E')} = t - t_{obs}.
\end{equation}
The $b(E)$ term is the energy loss rate, ${\bf r_s}$ indicates
the source position, and $\lambda$ is the typical propagation length defined as:
\begin{equation}\label{eq:lambda}
    \lambda^2 = \lambda^2(E,E_s) \equiv 4\int_E^{E_s} dE' \frac{D(E')}{b(E')} 
\end{equation}
where $D(E)$ is the diffusion coefficient taken as a double broke power-law in energy \cite{Di_Mauro_2023}. The $e^{\pm}$ energy losses include ICS off the ISRF and the synchrotron emission on the Galactic magnetic field.
The flux of $e^\pm$ at the Earth for a source of age $T$ and distance $d=|{\bf r_{\odot} - r_s}|$ is given by:
\begin{equation}
 \Phi_{e^\pm}(E) = \frac{c}{4\pi} \mathcal{N}_e(E,{\bf r=r_{\odot}},t=T).
 \label{eq:flux}
\end{equation}
 The propagation parameters are fixed to the ones derived in \cite{Di_Mauro_2023} from a fit to the latest data measured by the AMS-02 experiment, particularly the absolute fluxes of protons, He, C, O, N, B/C, Be/C, and Li/C after 7 years of data collection, from 2011 to 2018 \cite{AMS:2021nhj}, complemented with proton and helium data from Voyager \cite{stonevoyager2013}. 
 Specifically, we consider their benchmark model {\tt Conv $dv_c/dz$}, with the value of $L$ fixed at $4$ kpc (see Table 2 of \cite{Di_Mauro_2023} for the values of the parameters of this model). 
 We note that, however, the  parameter $L$ is not relevant in this study since we implement solutions without boundaries both in the radial and the vertical directions reported in Eq. \ref{eq:numberdensity}. The infinite halo approximation has been widely used to compute the flux from single sources located in the Galactic plane in~\cite{Fornieri_2020,Recchia_2019,Manconi:2020ipm}. We verified that our conclusions remain unchanged when adopting a more refined approach, incorporating the halo size $L = 4$~kpc using the image charge method described in \cite{1979ApJ...228..297C}, which modifies the infinite halo solution reported in our Eq.~\ref{eq:numberdensity} (see Eq. 2 of their paper). By evaluating 100 realizations of one of our setups, later referred to as the \textbf{$\bf \tau_0$ distribution} model (see Sec. \ref{sec:simulations}), we found that the mean deviation between the total $e^+$ flux from pulsars obtained with the boundary condition and the infinite halo approximation is less than 1\% at $E = 10$ GeV and decreases further at higher energies.

Energy losses are computed on the interstellar photon populations at different wavelengths following the default \textsc{Galprop} model, which is consistent with more recent estimates in the few kpc around the Earth \cite{Vernetto:2016alq}, taking into account Klein-Nishina corrections for ICS, and setting the Galactic magnetic field with an intensity of $B=3$ $\mu$G.

We also test how the results change considering a two-zone diffusive zone around each pulsar. Recent results \cite{Abeysekara:2017science,DiMauro:2019yvh,LHAASO:2021crt} suggest that the diffusion coefficient around Geminga, Monogem, and PSR J0622+3749  is $\sim 10^{26}$~cm$^2$/s at 1 GeV, {\it i.e.}~about two orders of magnitude smaller than the value derived for the entire Galaxy through a fit to AMS-02 CR nuclei data \cite{Kappl:2015bqa,Genolini:2015cta,Genolini:2019ewc}, although different interpretations have been proposed \cite{Recchia:2021kty}. 
A phenomenological description for this discrepancy proposes a two-zone diffusion model, where the region of low diffusion is contained around the source and delimited by an empirical radius $r_b$ \cite{Profumo_2018, Tang:2018wyr,Osipov_2020}.
We stress here that our main purpose is to derive the consequences of the presence of such inefficient diffusion zones around Galactic pulsars using such a phenomenological description, while no attempt is made to provide a detailed theoretical interpretation of this phenomenon (see for different possible explanations \cite{Recchia:2021kty,Evoli-Morlino-2018PhRvD..98f3017E,fang+19,mukhopadhyay+22,gupta+21,schroer+22,Plotnikov_2024}).
When using a two-zone diffusion scenario, we implement the solutions to the propagation equation reported in Ref. \cite{Tang:2018wyr,DiMauro:2019yvh}.
According to the results of \cite{Abeysekara:2017science,DiMauro:2019yvh,DiMauro:2019hwn,Schroer_2023}, the radius $r_b$ of the low-diffusion zones is at least $r_b>30$~pc. We will adopt a value of $r_b$=60 pc (see e.g. \cite{Manconi:2020ipm} for a discussion about the variation of $r_b$).

Finally, we estimate the $e^+$ dipole anisotropy that arises from each realization of Galactic pulsars. 
We note that to properly model a possible anisotropy, a complete spherical harmonic decomposition of the flux should be employed. In such a framework, the flux intensity can be expressed as a sum of multipoles depending on the direction: $I(\theta, \phi) = \sum A_{lm} Y_{l}^{m}(\theta, \phi)$, 
where $ Y_{l}^{m}(\theta, \phi)$ are the spherical harmonics and $A_{lm}$ are the constant coefficients. However, under the assumption that one or a few nearby sources dominate, the dipole term will likely provide the main contribution to the multipole expansion.
We here follow the formalism originally derived by \cite{GINZBURG1964XI} and summarized in  \cite{1971ApL.....9..169S,Manconi:2016byt} in order to compute the total $e^+$ anisotropy 
resulting from a collection of sources with specific positions in the sky, as provided in the ATNF catalog. 
We will end up with a prediction of the dipole anisotropy as a function of energy. 

The analysis is based on the standard, homogeneous diffusion model of CR propagation, without accounting for additional uncertainties that may arise from more refined propagation models. This standard model is calibrated to reproduce observations of diffuse CR nuclei fluxes but does not include localized effects near Earth. Such effects could be especially relevant for $e^+$ and $e^-$ fluxes, 
given their relatively shorter diffusion-loss length with respect to stable CR nuclei. 
In particular, this approach overlooks the influence of nearby structures like the Local Bubble \cite{Donato_2002}, superbubbles, and the orientation of the regular Galactic magnetic field, which can induce anisotropic diffusion. Models of the Galactic magnetic field, such as those described in \cite{Giacinti_2018,Beck_2016,Jansson_2012}, suggest that anisotropic diffusion can significantly alter the contribution of individual sources to the observed flux at Earth. While these considerations are beyond the scope of the current study, they underscore the interest to account for such effects, characterized by large uncertainties, in future analyses to achieve a more complete understanding of CR propagation and $e^\pm$ fluxes.

\subsection{Summary of setups and fit to AMS-02 data}
\label{sec:simulations}

\begin{table*}[t]
\centering
\begin{tabular}{ c @{\hspace{10px}} | c @{\hspace{10px}} | c @{\hspace{10px}}  |c @{\hspace{10px}}  } \hline\hline
\textbf{Pulsar} 
&   \textbf{Simulated }                        &  \textbf{ Benchmark  }  
&    \textbf{ Variations }  \\ 
\textbf{property} 
&   \textbf{quantity }                        &    
&    \textbf{ }  \\ 
\hline
Spin-down  & $\tau_0$ &  Distribution from \cite{Chakraborty:2020lbu} &   10 kyr\\
\hline 
$e^\pm$ injection & $\gamma_{\rm L}$ & Uniform [1.0-2.0]   & - \\ 
 & $\gamma_{\rm H}$ & Uniform [2.0-2.8]   & - \\ 
& $E_b$ & Uniform [300-600] GeV  & - \\ 
 &  $\eta$ &  Uniform [0.01-0.1]  & - \\ 
\hline 
Kick velocity & $v_k$ & -   & \texttt{FK06VB} \cite{2006ApJ...643..332F}  \\  \hline\hline
\end{tabular}
\caption{Summary of the emission parameters simulated for each catalog pulsar. The simulated quantities (first two columns), the distributions adopted in their simulation, together with the boundary of their validity range, for our benchmark case (third column), as well as the tested 
variations (last column) are reported.
See Section~\ref{sec:simulations} for details.}
\label{tab::models}
\end{table*}

As explained in the previous Sections, in order to compute the $e^+$ flux at the Earth for each known catalog pulsar, the source term  $Q(E,t)$ (see eq.~\eqref{eq:spectrum}) needs to be specified. This includes quantities that are not measured and provided directly in catalogs, for which we set up simulations,  while distance, age and $\dot{E}$ are taken directly from the catalog. 
Specifically, in our benchmark setup the quantities that we simulate are: $\tau_0$, $\gamma_{\rm L}$, $\gamma_{\rm H}$, $E_b$ and $\eta$. For an alternative setup, we also simulate the birth-kick velocity 
$v_k$. A summary of these simulated quantities, together with the distribution adopted to simulate each value is illustrated in Table~\ref{tab::models}. 

We recap and label here the  different setups investigated using AMS-02 data in what follows. 
For each realization we compute the $e^+$ flux from every source in the catalog sample defined in Section \ref{sec:atnf}. 

\paragraph{$\tau_0$ distribution.} 
 This is our benchmark setup. The $\tau_0$ values for each source are simulated following \texttt{CB20} \cite{Chakraborty:2020lbu}. $\eta$, $\gamma_{\rm L}$, $\gamma_{\rm H}$ and $E_b$ are extracted from uniform distributions, while the propagation in the Galaxy is modeled using the transport parameters corresponding to the model {\tt Conv $dv_c/dz$} from \cite{Di_Mauro_2023}.
\paragraph{$\tau_0$ fixed.} 
 Same as the benchmark setup, but with the value of $\tau_0$ fixed to 10 kyr for each source. 
\paragraph{Delayed injection.} 
Same as the benchmark setup, but considering only the $e^{\pm}$ emitted after the escaping of pulsars from the SNR. An additional parameter, the birth kick velocity, is simulated for each pulsar, and values are sampled adopting the distribution \texttt{FK06VB} reported in ~\cite{2006ApJ...643..332F}.
\paragraph{Two zone propagation} 
Same as the benchmark setup, but considering the presence of a slow diffusion halo around each pulsar with size $r_b=60$ pc and a diffusion coefficient $D(E) = 10^{26}$ cm$^2$/s for $r<r_b$.

\medskip 

Depending on the setup, a number ranging between 1000 to 3000 realization $N_{rea}$ of the $e^+$ emission parameters is performed, as some models adapt more easily to the data than others. The $e^+$ flux from each Galactic pulsar is computed accordingly. Each realization consists of the sum of the contribution of all the Galactic pulsar considered, and is fitted to AMS-02 $e^+$ data \cite{PhysRevLett.122.041102} above 10 GeV, to minimize the impact of solar modulation and other low-energy effects \cite{Boudaud:2016jvj}.
In each fit, a secondary component coming from cosmic ray fragmentation in the ISM is added, consistently  with the implemented propagation model, as derived in \cite{Di_Mauro_2023}. Both the total contributions of pulsars and the secondary component are multiplied by normalization factors, $A_P$ and $A_S$ respectively, whose values are determined for each simulation through the fitting procedure.
The secondary component's normalization factor, $A_S$, is allowed to vary between 0.01 and 2. The broad range considered here is justified by the fact that secondary production is significantly influenced by the diffusion coefficient associated with various sizes of the diffusive halo, denoted as $L$. On the other hand, in the energy range under examination (10 GeV$-$1 TeV), where energy losses dominate, the contribution from primary sources does not exhibit this pronounced dependency on the diffusion coefficient, and its sensitivity to changes in $L$ is quite weak. Adjusting the normalization of the secondary production provides an effective means of exploring potential values for $L$ and, consequently, the diffusion coefficient, without the need for recalculating the primary flux.

Similarly to $A_S$, the total flux from all pulsars is adjusted by an overall normalization factor, $A_P$.  Both are determined for each realization through the fitting procedure.

Predictions are corrected for solar modulation using the force field approximation, with the Fisk potential $\phi$ free to vary between 0.4 and 1.2 GV. The comparison with AMS-02 $e^+$ data is performed via standard $\chi^2$ minimization. We disregard systematic error correlations in the AMS-02 data due to the lack of information provided by the Collaboration~\cite{PhysRevLett.122.041102}.

For each setup, the realizations are filtered using as criteria the quality of the fit of the cumulative emission from all the Galactic pulsars and the secondaries to AMS-02 data. The contribution from each individual source is stored to identify a possible set of sources contributing significantly to the $e^+$ flux, and their underlying properties.

\section{Results}\label{sec:results}
The aim of this paper is to probe if the $e^+$ flux measured by the AMS-02 experiment \cite{PhysRevLett.122.041102} can be explained merely by a secondary component and by the emission from catalogued pulsars. In those models where this occurs, we investigate a  hierarchy among the most relevant emitters.

\begin{table}[t]
\begin{center}
\begin{tabular}{ |c|c|c|c|c|c| } 
 \hline
         & $N_{rea}$ & $\chi^2_{\rm red}<1.5$  & $A_S$ & $A_P$ & $N_{10^{-4}}$ \\
 \hline
 \textbf{$\mathbf{\tau_0}$ distribution} &          3000   &        72       & 1.3 &  0.6 & 3.6\\      
 \textbf{$\mathbf{\tau_0}$ fixed} &          1000     &        768       & 1.0 &  0.8 & 6.7 \\      
 \textbf{Delayed injection} &          2000    &        194       & 1.2  &  4.6  & 5.3 \\      
 \textbf{Two zone} &          3000    &        159       & 1.5 & 0.2    & 3.1 \\

 \hline
\end{tabular}
\caption{
 In order: model considered; number of total realizations $N_{rea}$ tested for each model; number of realizations that produce a  $\chi^2_{\rm red}$ smaller than 1.5 in the fit to AMS-02 data \cite{PhysRevLett.122.041102}; average normalization of the secondary flux ($A_S$) and of the total pulsar contribution ($A_P$) obtained for the simulations that produce a $\chi^2_{\rm red}<1.5$; average number of sources satisfying the $10^{-4}$ criterion (see the text for details) for the simulations that produce a $\chi^2_{\rm red}<1.5$.}
\label{tab:simulation_result}
\end{center}
\end{table}

\subsection{Results from fits to AMS-02 $e^+$ data}
\label{sec:results_fit}

As a first result, we show in Table~\ref{tab:simulation_result} the number of realizations that provide a good fit to AMS-02 data, having $\chi^2/d.o.f. = \chi^2_{\rm red}<1.5$, and for each setup (see Sect. \ref{sec:simulations}).
First of all, we note that each framework finds realizations which fit well the AMS-02 data.
Notwithstanding that the statistical variation of their number does not bring a deep physical meaning, the \textbf{$\mathbf{\tau_0}$ fixed} model fits nicely the data most of the times. 
A fixed $\mathbf{\tau_0}$ for all the pulsars, clearly unrealistic, intrinsically predicts similar contributions among the sources (see Eq.~\eqref{eq:W0PWN}). It is then 
more likely to have a smoother total contribution to the $e^+$ flux than in 
cases in which few very bright sources can be switched on, and thus an easier fit to the pretty unstructured data. This also means that the other simulated parameters do not significantly impact the $e^+$ production, leading to realizations that are quite similar to each other.
At variance, the \textbf{$\mathbf{\tau_0}$ distribution} modeling can give rise to  potentially extremely bright sources during the early stages of their life. 
The resulting total $e^+$ flux from all the pulsars can thus have peculiar structures, which harshly fit the data. Regarding the allowed overall normalization $A_P$ of the pulsar primary flux for \textbf{$\mathbf{\tau_0}$ fixed} and \textbf{$\mathbf{\tau_0}$ distribution}, we generally find values slightly smaller than one, compatible with previous works \cite{Cholis_2018, Orusa_2021,Cholis:2022kio}.

\begin{figure*}[t]
    \begin{center}
    \includegraphics[width=0.49\textwidth]{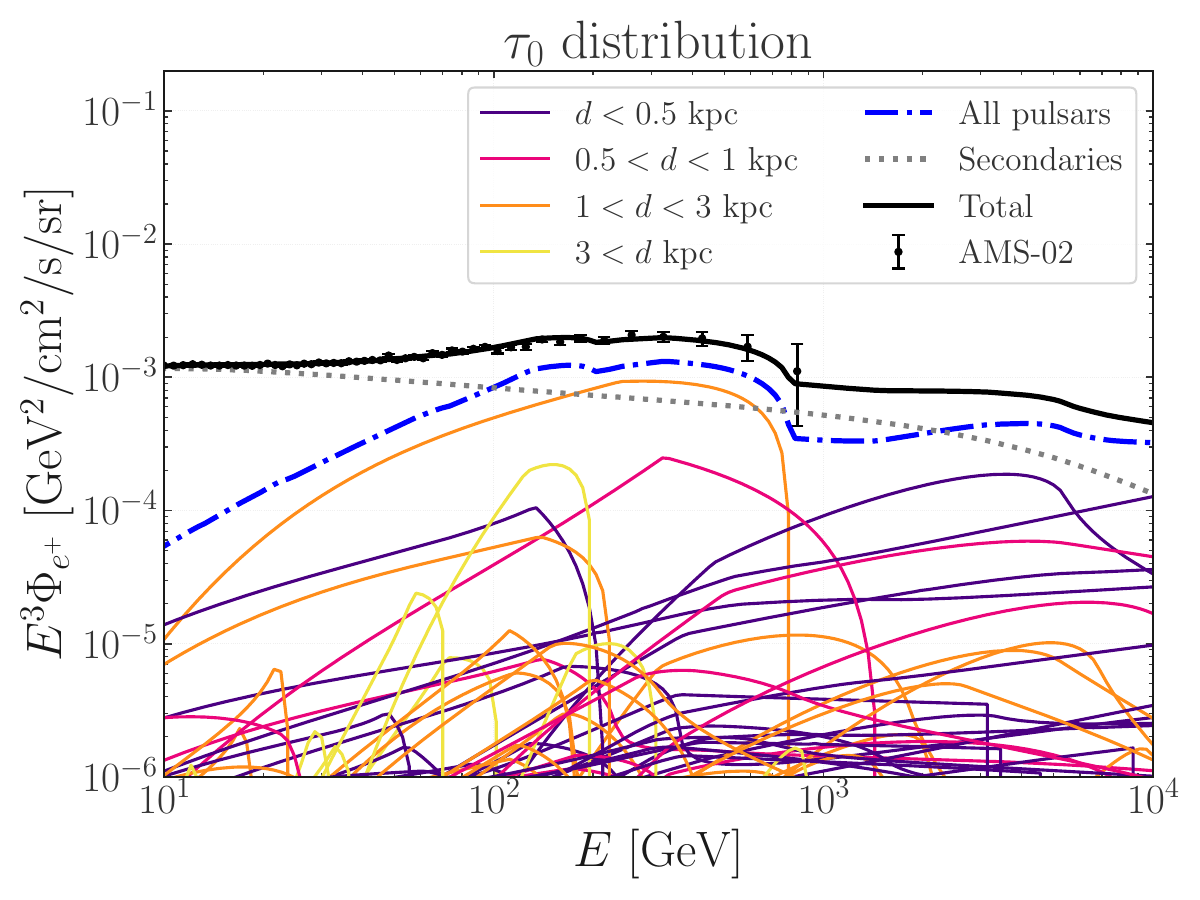}
    \includegraphics[width=0.49\textwidth]{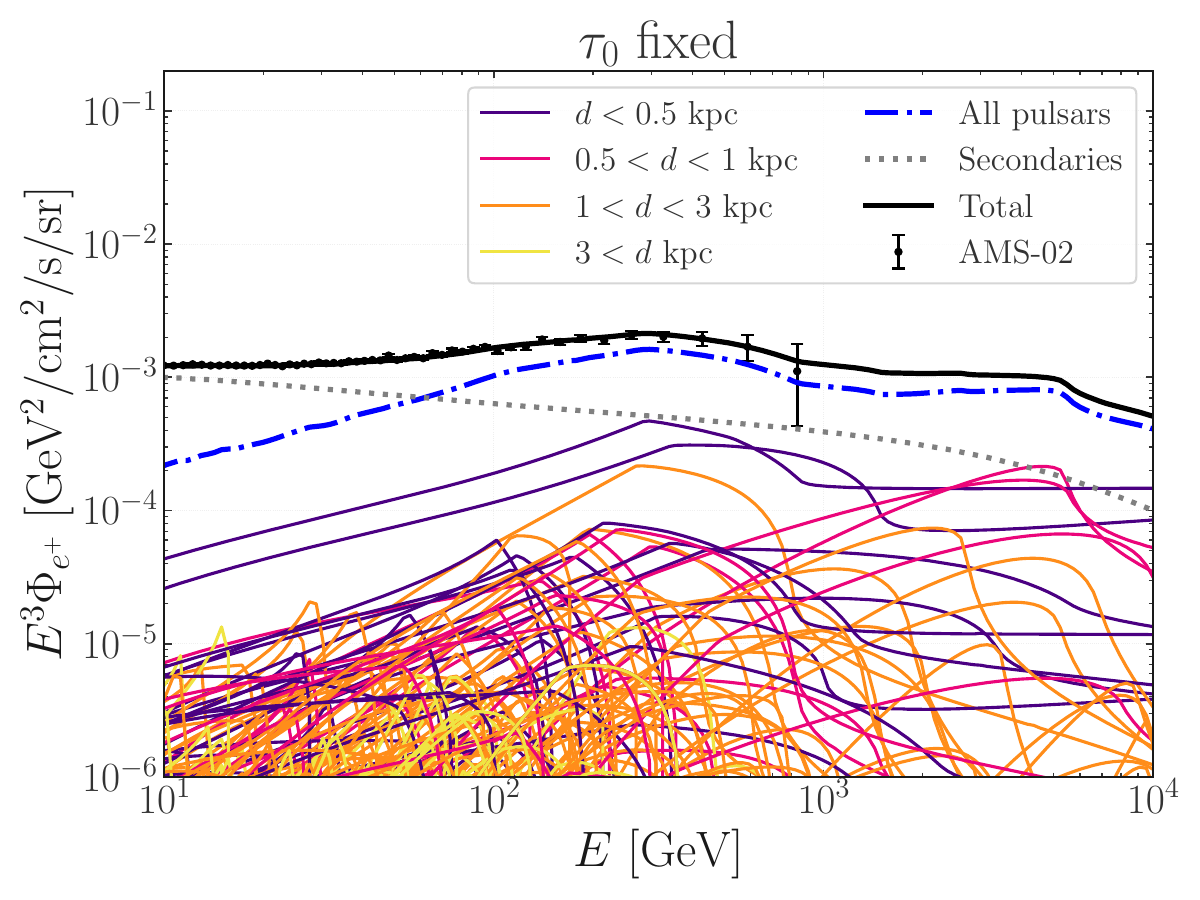}
    \includegraphics[width=0.49\textwidth]{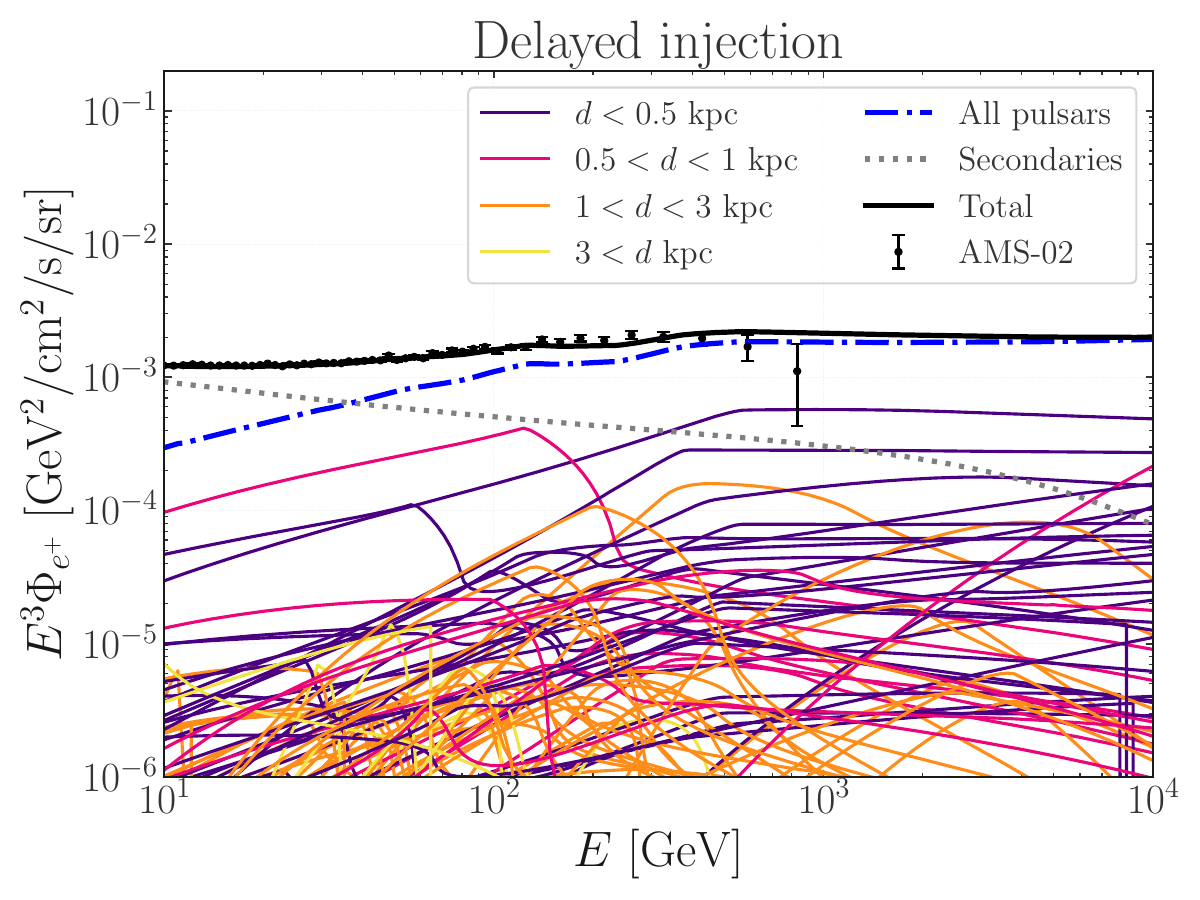}
    \includegraphics[width=0.49\textwidth]{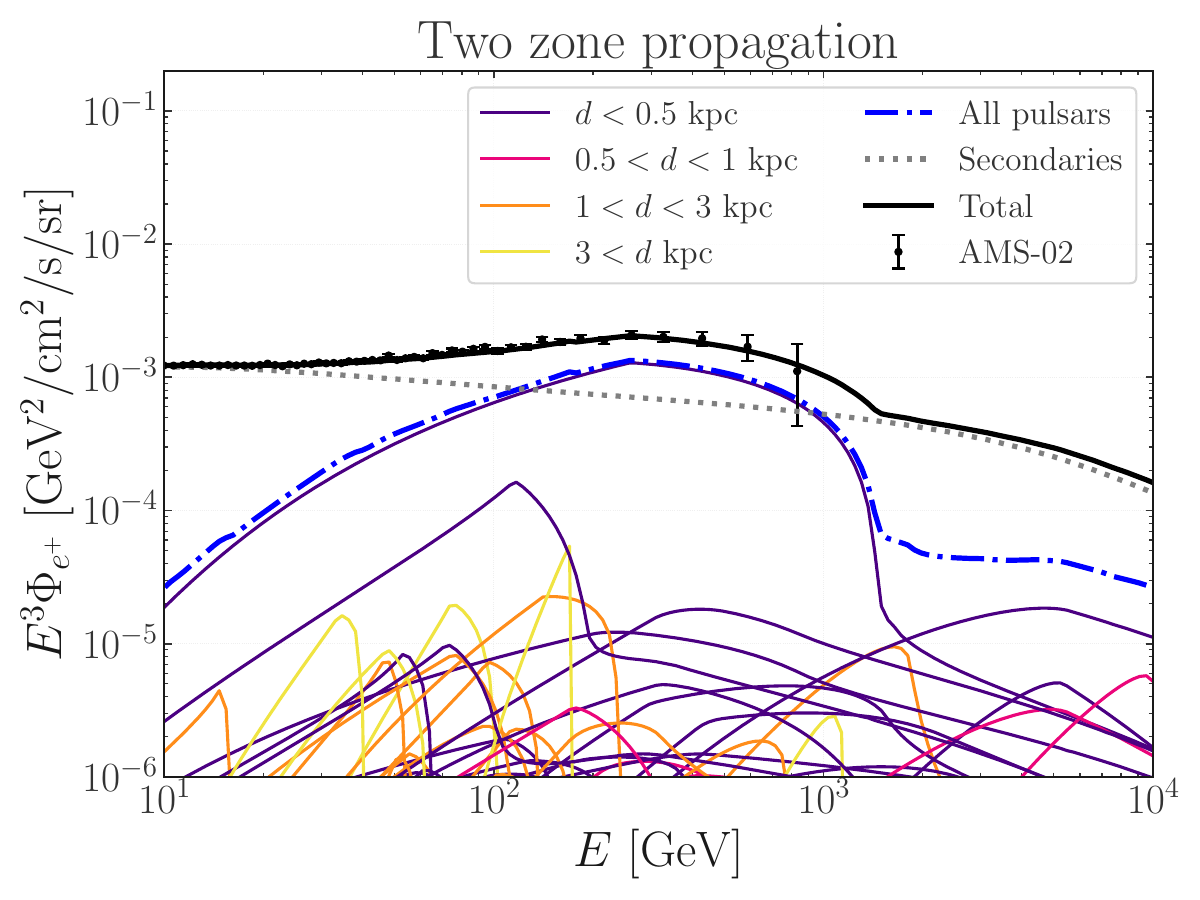}
    \caption{Comparison between the AMS-02 $e^+$ flux data \cite{PhysRevLett.122.041102} (black points) and the  flux  from secondary production (grey dashed line) and pulsars (blue dashed line) for four realizations of the different models that we tested with $\chi^2_{\rm red}<1$. The contributions from each source, reported with different colors depending on their distance from the Earth, are shown.} 
    \label{Fig:fluxes_simulation}
    \end{center}
\end{figure*}

In the \textbf{Delayed injection} setup, only the $e^+$ produced after the pulsar has left the remnant are considered. This approach reduces the influence of extremely bright sources, as most $e^+$ are produced in the early stages of the pulsar's lifetime, i.e. $\lesssim 50$ kyr. Consequently, the total flux 
has not strong peaks and follows generally a smooth trend. 
However, the fit requires efficiencies between 5 and 46\% - implied in the value of $A_P$ - since the model does
not account for the significant amount of $e^+$ produced at the beginning of the pulsar's lifetime. Efficiencies close to 50\%, although compatible with previous estimates obtained with the same model \cite{Orusa_2021,Evoli_2021}, are in tension with the values obtained from the analysis of $\gamma$-ray halos around pulsars \cite{DiMauro:2019yvh,Recchia:2021kty,Schroer_2023,Amato:2024dss}, assuming a slow diffusion around this sources.  They result  although to be compatible with the ballistic scenario proposed in \cite{Recchia:2021kty}, which does not invoke a suppressed diffusion around pulsars. Changing the value of $E_1$ in the normalization of the spectrum from 0.1 to the $e^+$ mass can, for some specific set of parameters in the injection spectrum ($\gamma_L$ close to 2 and $\eta > 33$\% of the single pulsars), lead to unphysical values of $\eta$ for individual sources. However, this occurs only in a small region of the parameter space. The average value of $\eta$ required remains below 50\%.
In all the three models, good fits come with a normalization $A_S$ of secondary $e^+$ close to one, which marks a significant improvement compared to previous predictions \cite{Orusa_2021, DiMauro:2014iia}. We ascribe this result to the 
new computation of the secondary source term, in which the $e^+$ production cross sections have been adapted to a bunch of accelerator data \cite{Orusa_2022}.

For the \textbf{Two zone}, we obtained results similar to \textbf{$\mathbf{\tau_0}$ distribution}, although with a higher $A_S$. This outcome can be attributed to the physics of two-zone diffusion model, which tends to trap $e^+$ longer around the pulsar bubble, potentially preventing older sources from contributing to the $10-40$ GeV range. 
However, the goodness of the fits is somehow dimmed by the necessity of some
up-scaling of the secondary flux ($A_S=$1.5 on average) and a severe downgrading of the overall pulsar contribution ($A_P=$0.2 on average).

In Figure~\ref{Fig:fluxes_simulation}, we display the $e^+$ flux for four illustrative realizations with $\chi^2_{\rm red}<1$, across different setups. The plots include contributions from each catalogued pulsar, the secondary emission, and their sum alongside the AMS-02 data. 
Overall, the contributions from pulsars become significant above $10-20$ GeV, dominate in the $40-1000$ GeV range, and show variegated features at unconstrained energies above 1 TeV, depending on the specific realization. 
The most intense contributions come from sources located within 1 kpc. However, statistical variations can occasionally promote more distant sources to produce a large $e^+$ flux, as shown in the upper left panel of Fig. \ref{Fig:fluxes_simulation}.

\subsection{The most relevant ATNF pulsars}\label{sec:top10_pulsars}

\begin{figure*}[t]\centering
    \includegraphics[width=0.49\textwidth]{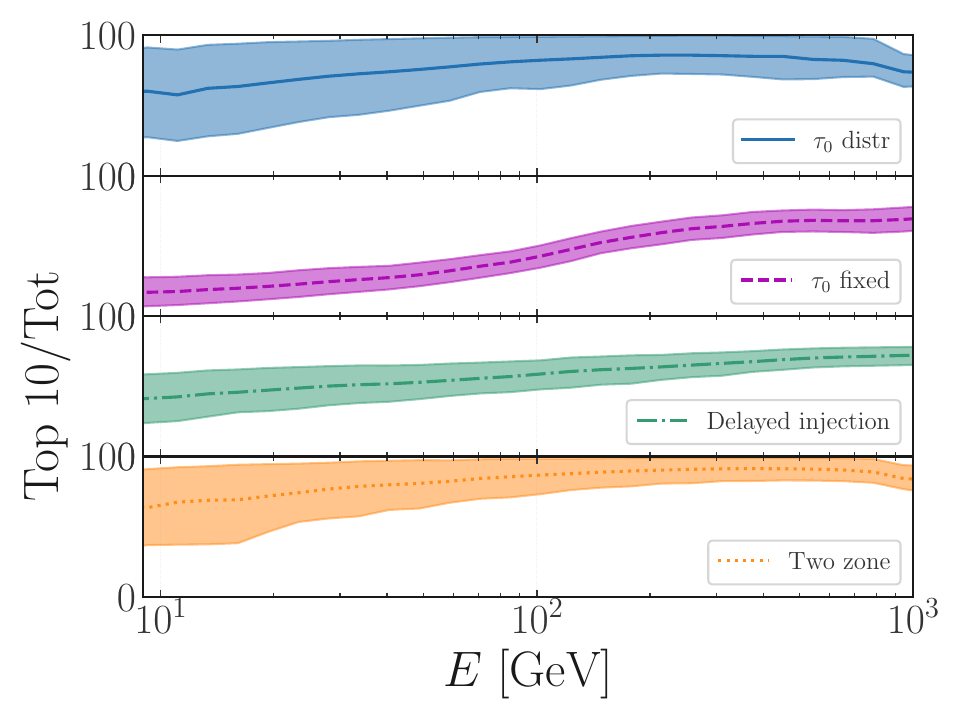}
    \includegraphics[width=0.49\textwidth]{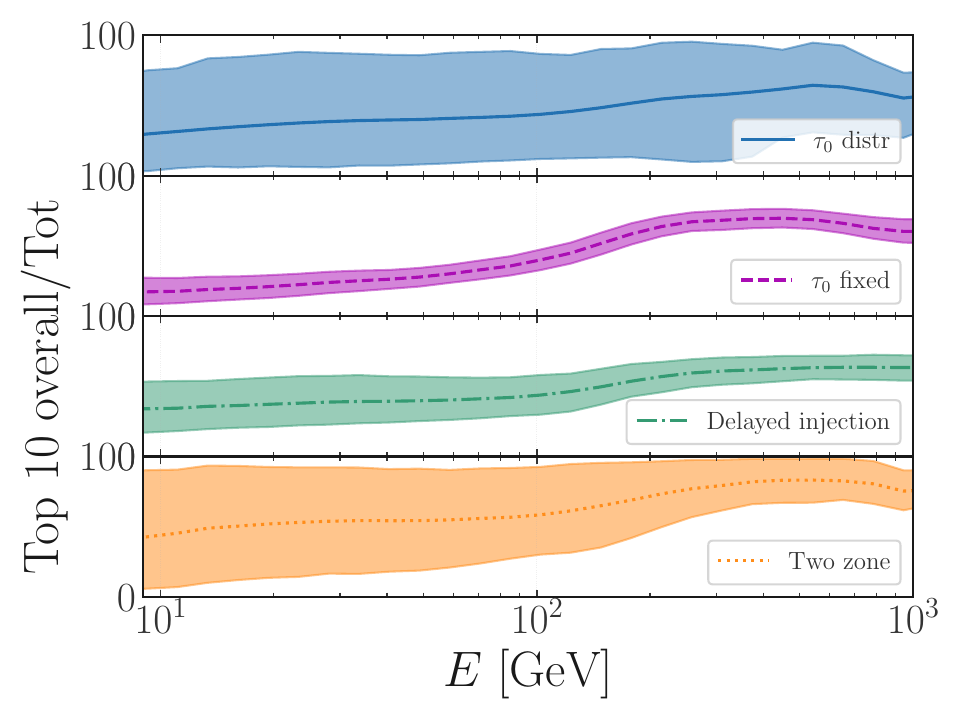}
    \caption{Left panel: the average flux percentage produced by the top 10 brightest sources in each realization that fits the data, along with the 68\% containment band, for the different models, relative to the total flux produced by all pulsars. Right panel: same as left panel, but considering this time  the top 10 brightest sources among the different models listed in column 5 of Table \ref{tab:table_pulsars}.} 
    \label{Fig:top10_percentage}
\end{figure*}

One of the main goals of this paper is to identify the sources that most significantly contribute to the $e^+$ flux and shape its profile, based on different models of pulsar evolution, and $e^\pm$ emission and diffusion. This will help prioritize the sources and provide indications for multi-messenger observations, in order to better understand their injection physics.

Our initial goal is to verify whether a small number of bright sources are sufficient to account for the majority of the $e^+$ excess, as found in our previous work based on full simulations of the Galactic pulsar population \cite{Orusa_2021}.
We show in Fig.~\ref{Fig:top10_percentage} (left panel) the average flux percentage produced by the top 10 brightest sources, in each realization that fits the data, along with the 68\% containment band for the different models, and relative to the total flux produced by all pulsars. Focusing on energies above 100 GeV, the top 10 pulsars contribute nearly 80\% of the total flux for all models.
In the \textbf{$\mathbf{\tau_0}$ distribution} and \textbf{Two zone} cases, which are more sensitive to the dominance of a few sources, the 10 brightest sources explain almost the total of the 
$e^+$ produced in the Galaxy. Meanwhile, \textbf{$\mathbf{\tau_0}$ fixed} and \textbf{Delayed injection} lower edge of the bands show a minimum contribution around 50\%.

\begin{figure*}[t]
    \includegraphics[width=0.49\textwidth]{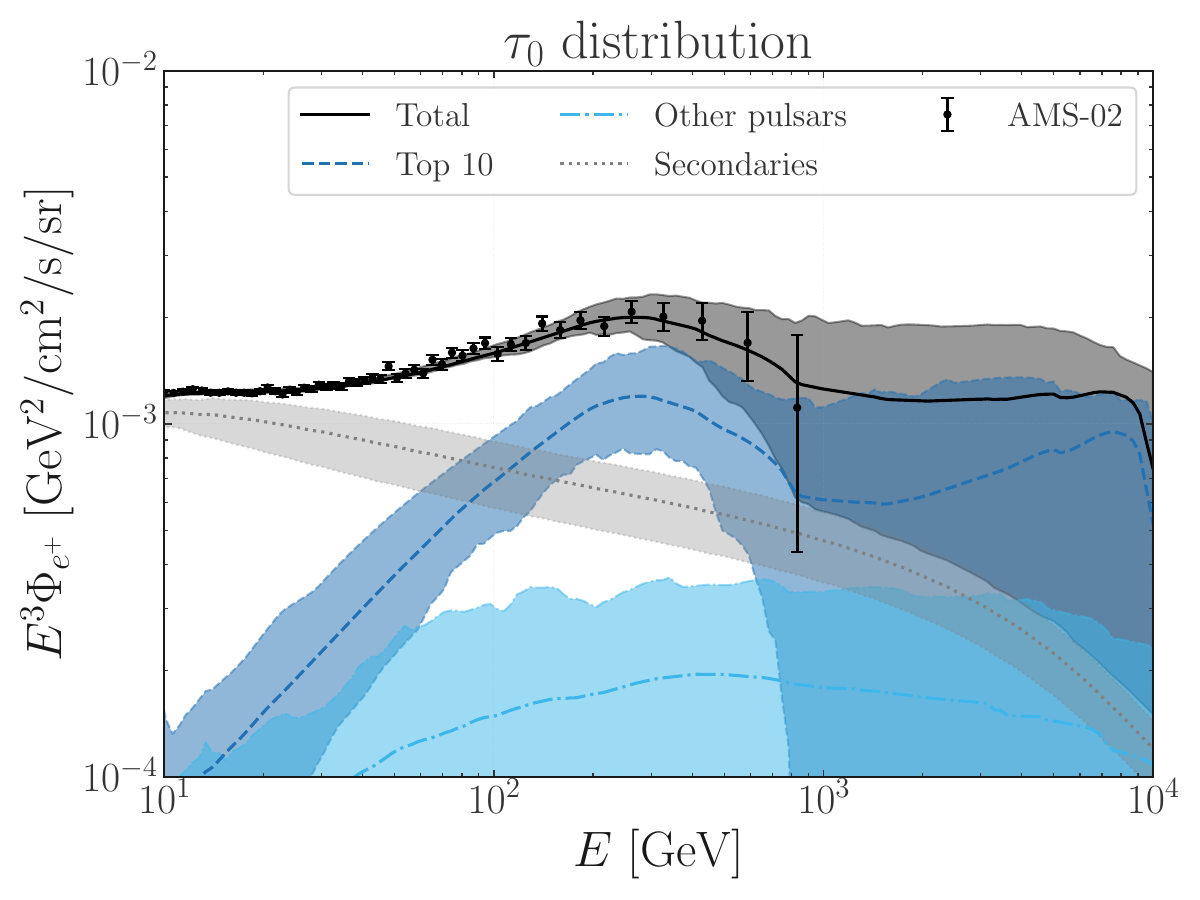}
    \includegraphics[width=0.49\textwidth]{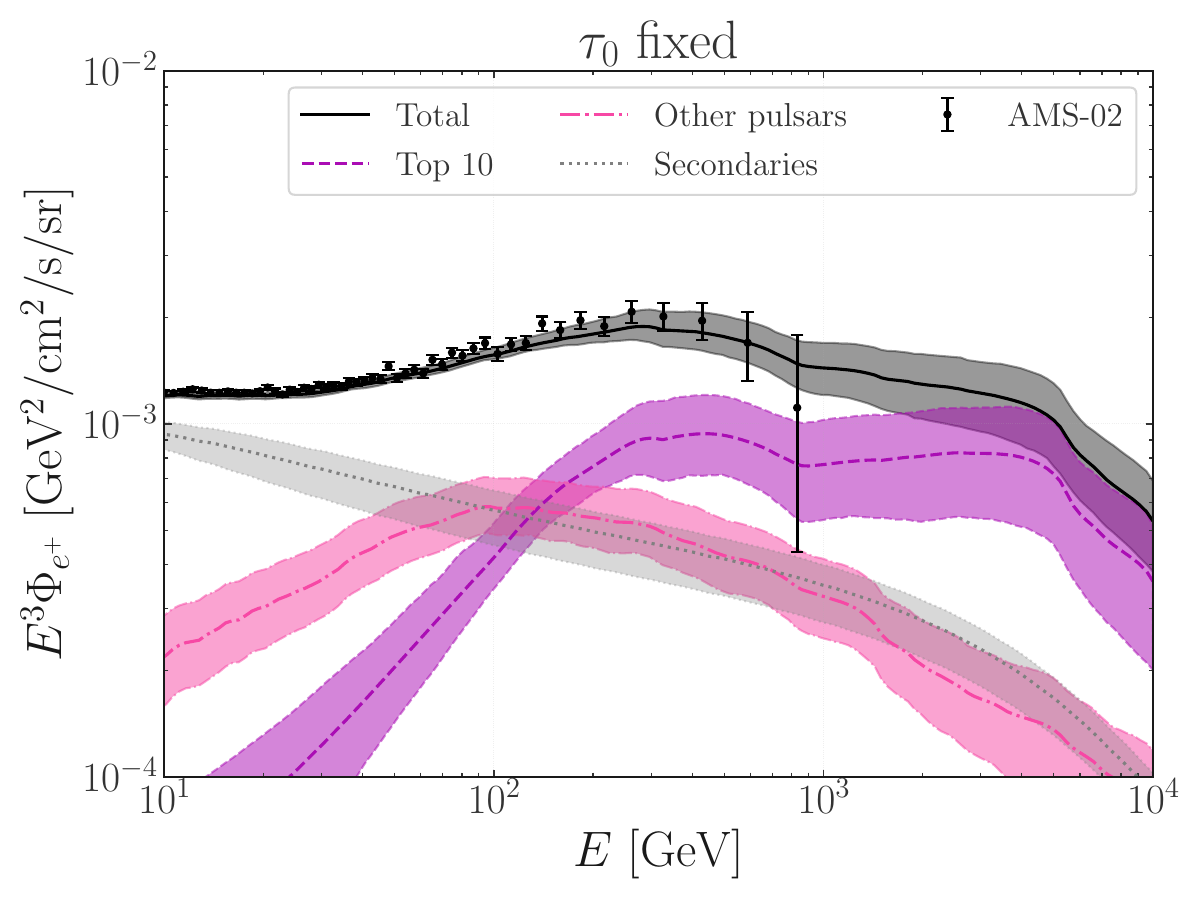}
    \includegraphics[width=0.49\textwidth]{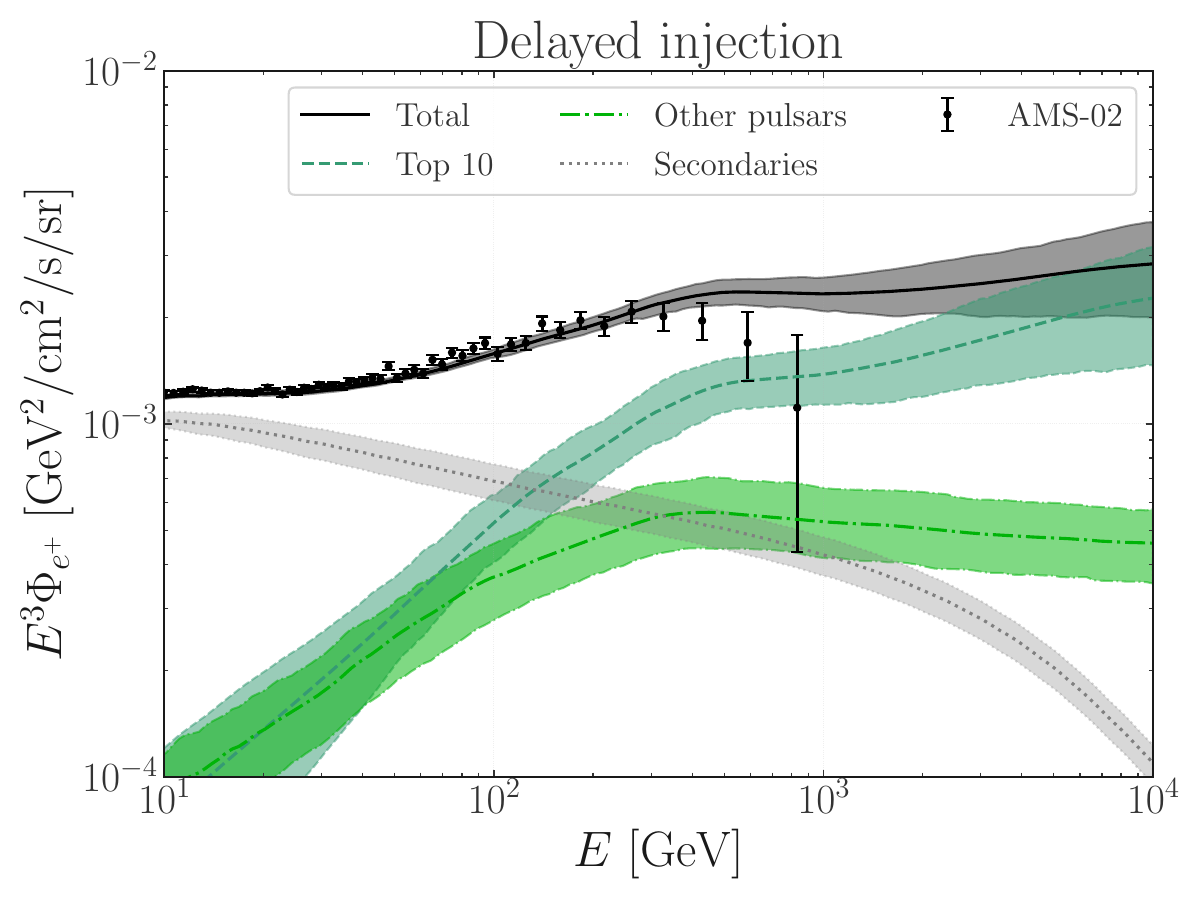}
    \includegraphics[width=0.49\textwidth]{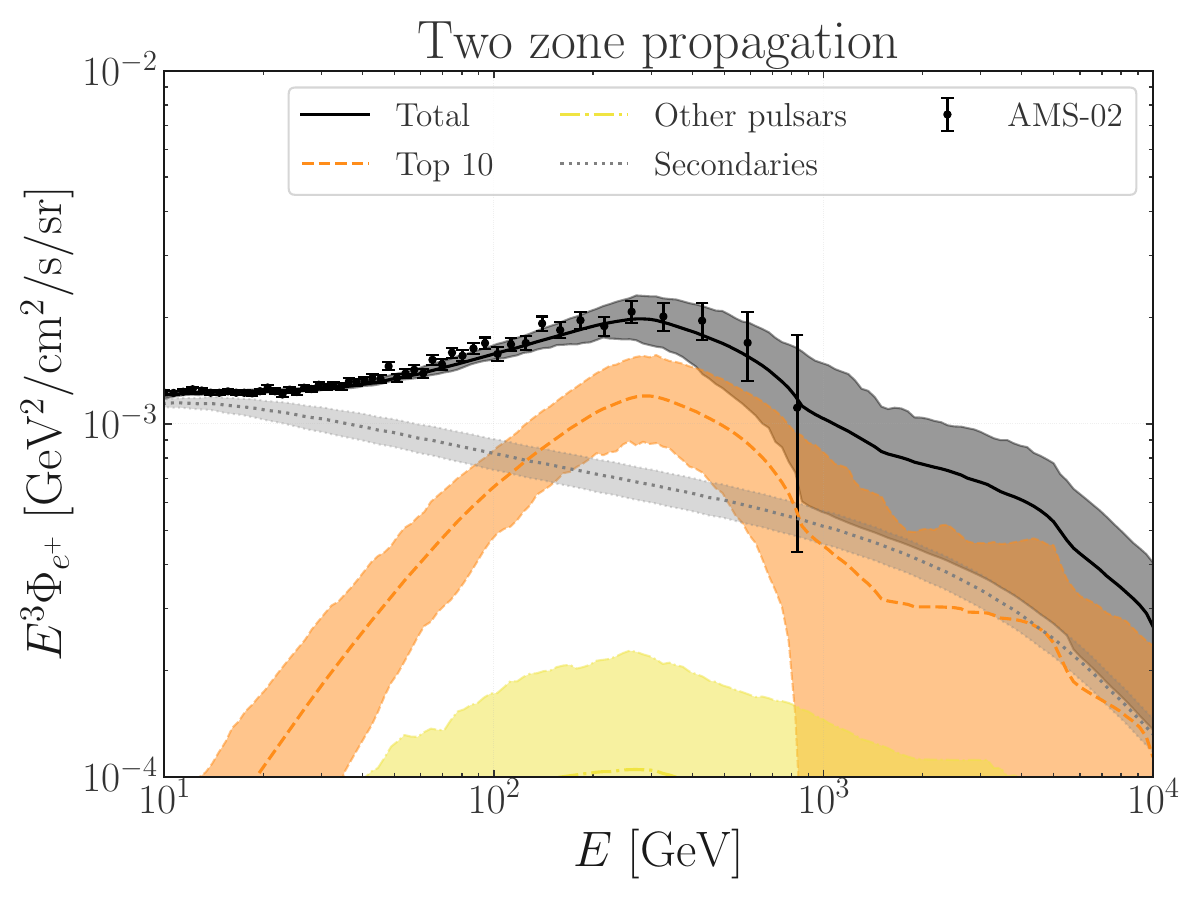}
    \caption{Average values of the fluxes of the top 10 brightest pulsars for each realization that successfully fit the AMS-02 $e^+$ data, of the secondary flux, of the total flux, and of the contributions from other pulsars, along with the 68\% containment band, for the 4 different models.} 
    \label{Fig:top_flux}
\end{figure*}

In Fig.~\ref{Fig:top_flux} the average values of the fluxes of the top 10 brightest pulsars are shown for each realization that fits the data, alongside the secondary flux, the contribution from the other pulsars, and their sum.
We also report the 68\% containment band. 
The role of the 10 brightest pulsars in shaping the high-energy total flux is evident everywhere. Also in the \textbf{$\mathbf{\tau_0}$ fixed} case, where very few dominant sources are not expected, the global trend is influenced by the top 10 pulsars.

We then apply a threshold criterion to determine which sources are responsible for the most significant contribution of pulsars to the $e^+$ emission, in realizations that fit the data. 
The rule  - referred to as the $10^{-4}$ criterion - is for a source to have a flux at an energy $E$ such that $E^3 \Phi (E)>10^{-4}$ GeV$^2/$cm$^2$/s/sr for at least one energy value between 10 GeV and 1 TeV.
The average number of sources satisfying this criterion across the different setups is reported in Column 6 of Table \ref{tab:simulation_result}. 
The number of powerful sources according to this criterion is low, between 3.1 and 6.7.
As expected, \textbf{$\mathbf{\tau_0}$ fixed} is the model with the highest number of contributing sources, justified by the adoption of the same $\tau_0$. 

\begin{figure*}[t]
\centering
\includegraphics[width=0.49\textwidth]{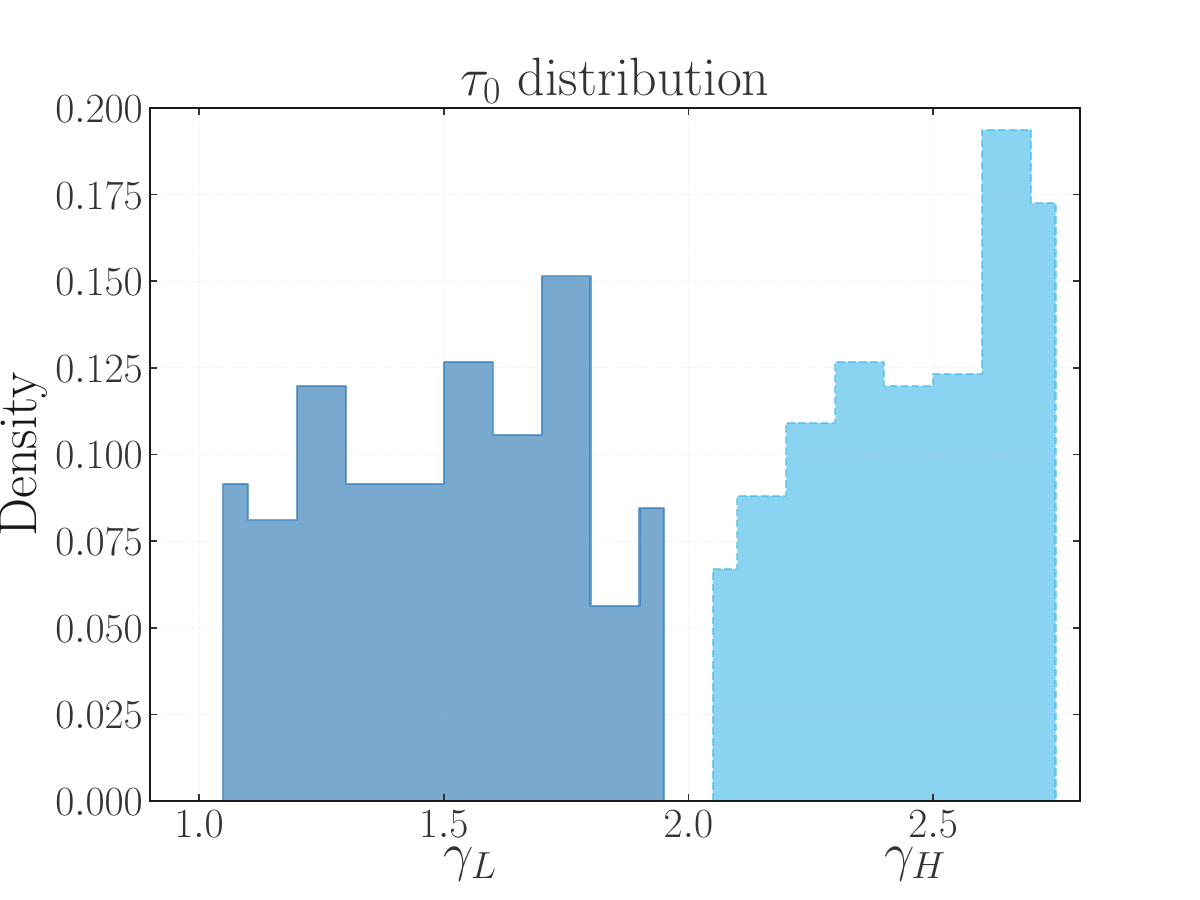}
\includegraphics[width=0.49\textwidth]{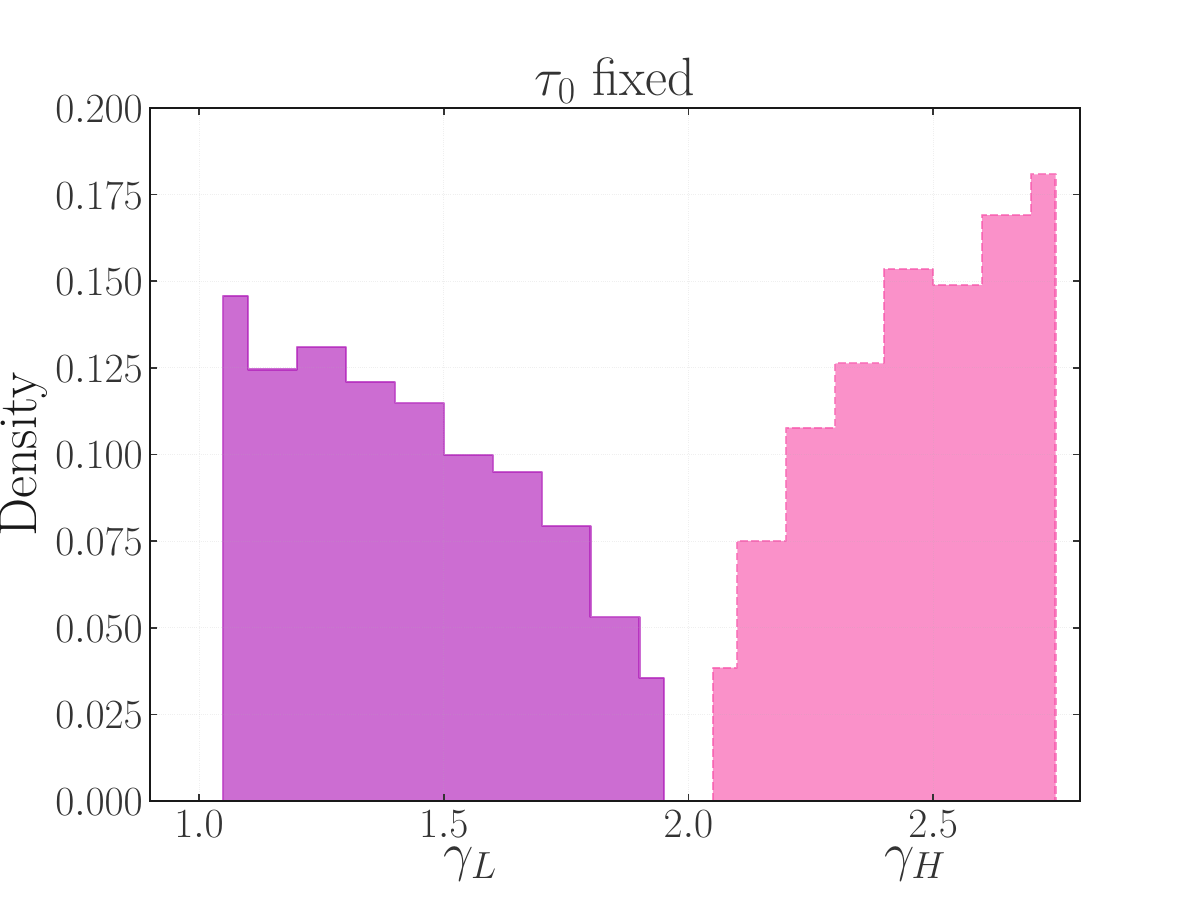}
\includegraphics[width=0.49\textwidth]{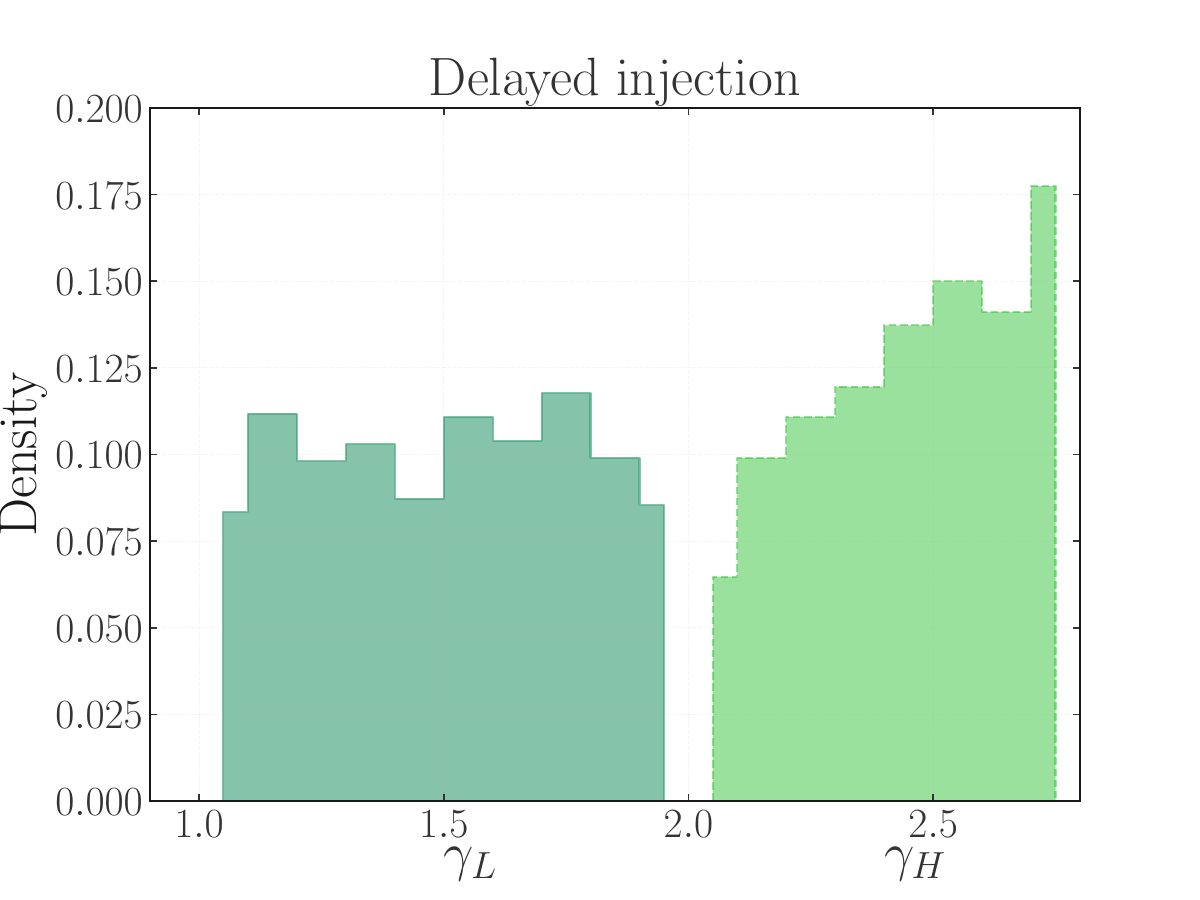}
\includegraphics[width=0.49\textwidth]{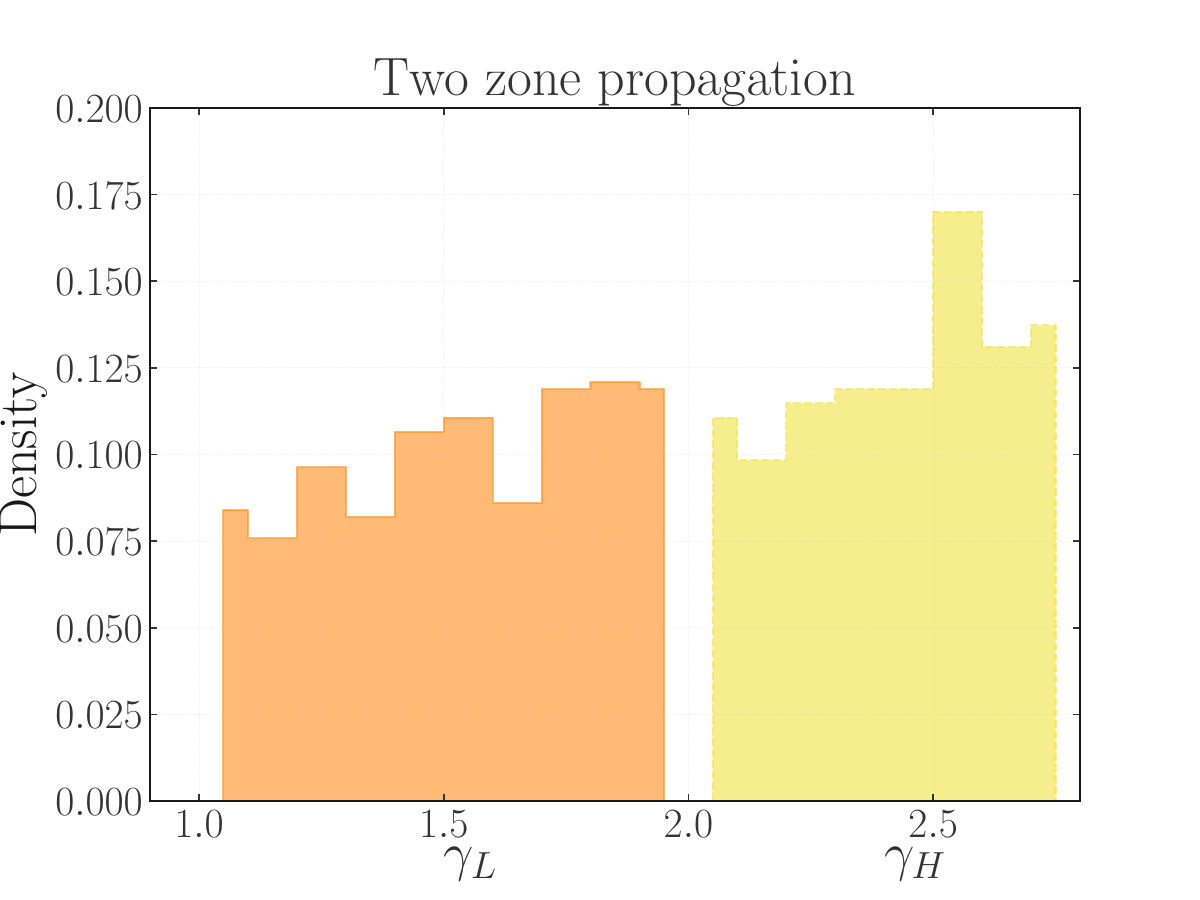}\caption{Distribution of the emission spectrum parameters for the brightest sources as selected from the realizations that successfully fit AMS-02 $e^+$ data for the four setups investigated: $\tau_0$ distribution (upper left), $\tau_0$ fixed (upper right), delayed injection (lower left), two zone propagation (lower right). In each panel, the histograms (normalized to the total number of selected realizations) of the two injection spectral parameters $\gamma_L$ and $\gamma_H$ as defined in Eq.\eqref{eq:spectrum} are reported.} 
    \label{Fig:emission_spectra}
\end{figure*}

We now investigate possible correlations and  patterns in the injection parameters of the brightest sources, considering only the realizations that successfully fit the data. Fig. \ref{Fig:emission_spectra} shows the distributions of $\gamma_L$ and $\gamma_H$ (see Eq.~\eqref{eq:spectrum}) for the different models, obtained for the bright sources able to satisfy the $10^{-4}$ criterion. A clear trend emerges, favoring high values of $\gamma_H$, a requirement previously highlighted in \cite{Evoli_2021}. This differs from what is typically inferred from the emission of well-studied PWNe, such as the Crab Nebula \cite{Kargaltsev_2008,Gelfand_2009,bucciantini+11,2008A&A...485..337V,Torres:2014iua,2015Sci...347..406H,2020A&A...640A..76P}, which favors values of $\gamma_H \sim 2.3-2.5$. This discrepancy may indicate that the particle population released by these objects undergoes significant energy losses before escaping the vicinity of the source, thereby softening the spectrum. On the other hand, we do not observe a clear trend in the $\gamma_L$ distribution, except in the model with a fixed $\tau_0$, which slightly favors smaller $\gamma_L$ values. This model, with its adoption of a fixed and relatively large $\tau_0$, tends to produce a fairly flat $E^3 \Phi (E)$ across all energies. To account for the rise of $e^+$ flux in the excess region, a hard spectrum before the break and a steep one afterward are required, allowing $E^3 \Phi (E)$ to rise for energies up to 300 GeV and then decrease for higher ones to match the data. Finally, the distribution of $E_b$ reflects the uniform sampling applied initially, with no clear trend emerging.

The $10^{-4}$ criterion serves also as a filter to highlight pulsars that consistently satisfy this condition across different models, allowing to establish a hierarchy of sources.
In Table~\ref{tab:table_pulsars}, left block, we report the names of the top 10 pulsars that satisfy more frequently the $10^{-4}$ criterion (always for the realizations that fits well the data), for all the different setups. In the right block we report the overall top 10 pulsars, determined by combining the frequency of occurrence across all models and averaging by the number of realizations that fit the data for each model, together with their distance, age and spin-down energy. The bold text style refers to the top 10 pulsars from the overall frequency study.

\begin{figure}[t]
    \begin{center}
    \includegraphics[width=0.49\textwidth]{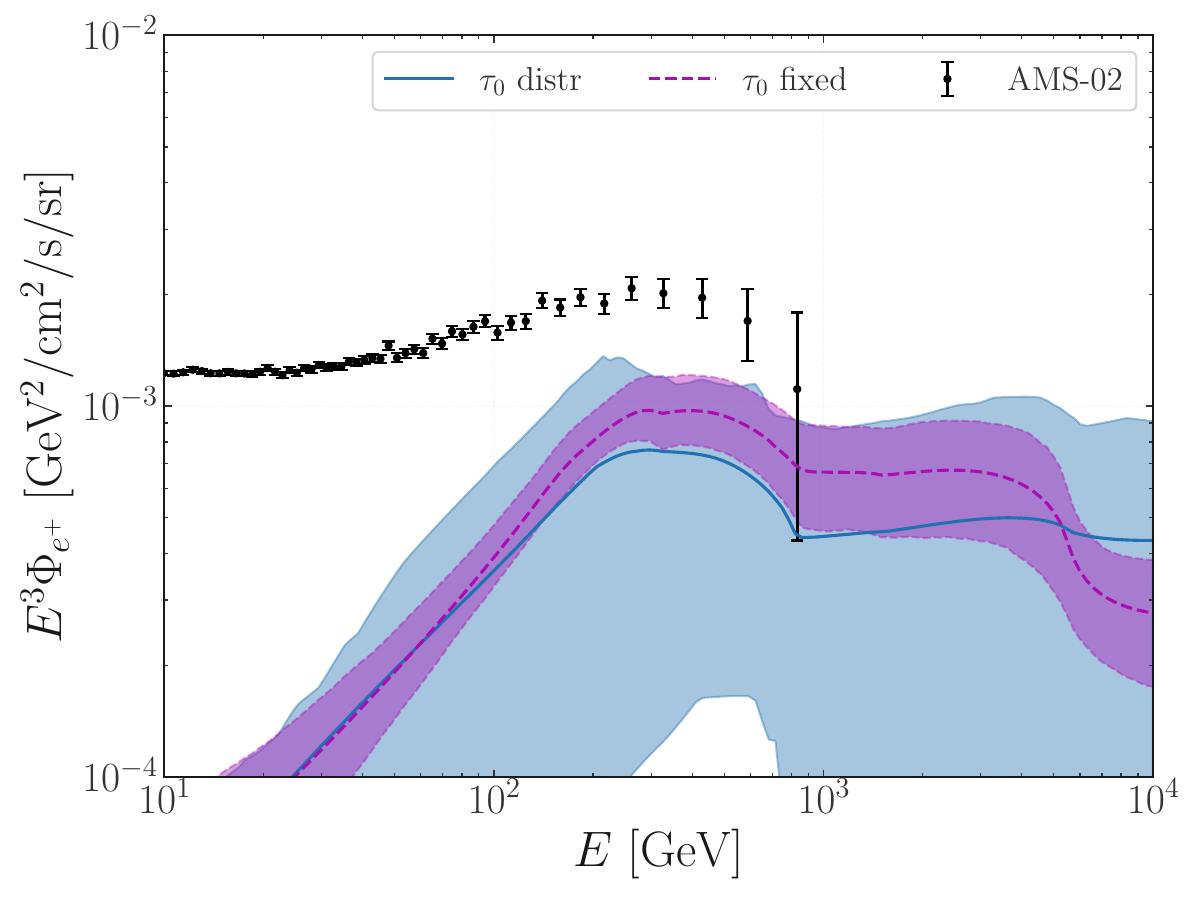}
      \includegraphics[width=0.49\textwidth]{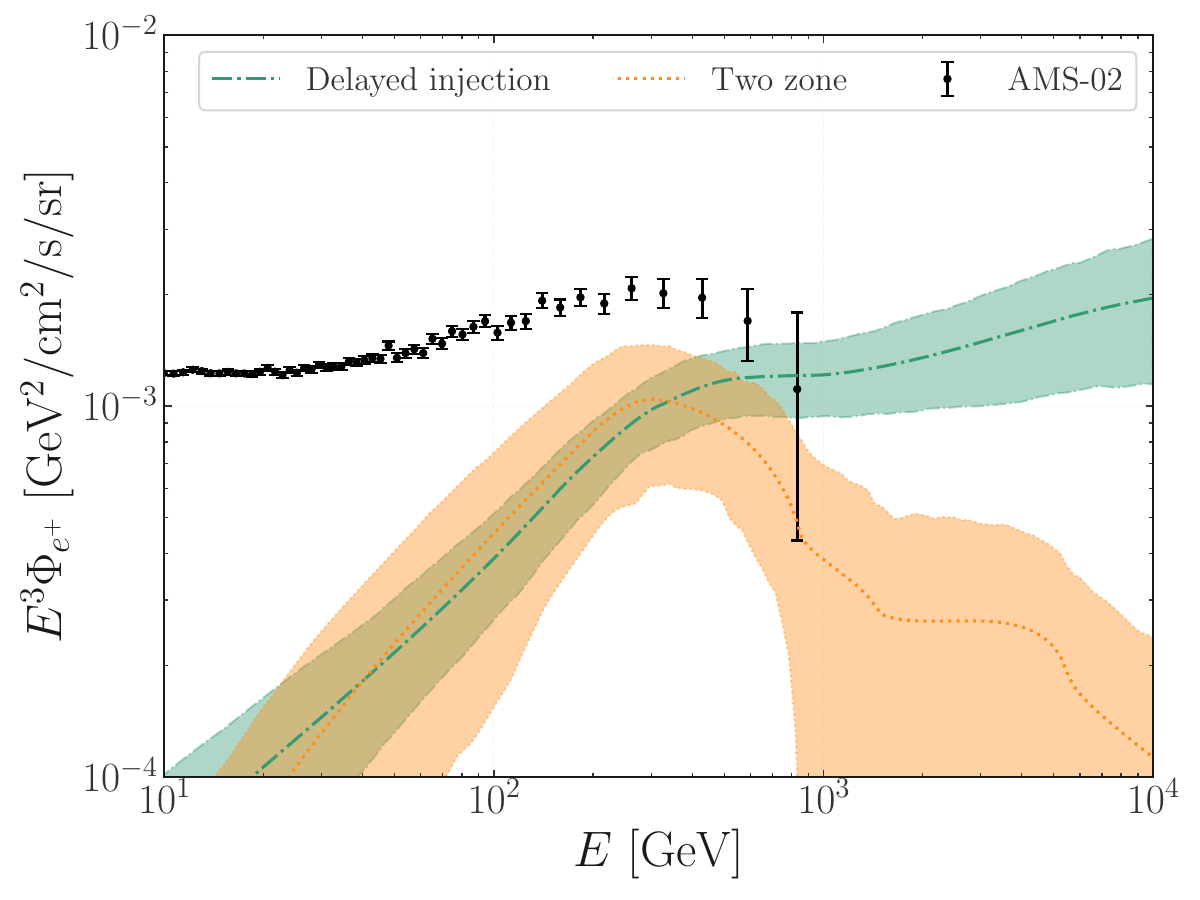}
    \caption{Flux of the top 10 overall sources identified with the $10^{-4}$ criterion and reported in column 5 of Table \ref{tab:table_pulsars}.} 
    \label{Fig:top10_flux}
    \end{center}
\end{figure}

In Fig.~\ref{Fig:top10_percentage} (right panel), the average flux percentage produced by the top 10 brightest sources listed in column 5, Table \ref{tab:table_pulsars}, is shown along with the 68\% containment band for the different models, normalized to the total flux produced by all the pulsars. With respect to the left panel of Fig.~\ref{Fig:top10_percentage}, for \textbf{$\mathbf{\tau_0}$ distribution} there is a larger uncertainty band, both due to the lower number of realizations that fit the data and to the high statistical variance that can promote one random source to be extremely bright when $\tau_0$ occurs to be small. The results for \textbf{$\mathbf{\tau_0}$ fixed} and \textbf{Delayed injection}, namely the models that smooth the most the statistical fluctuations, are similar to the left panel. 
In the \textbf{Two zone} model, the $e^+$ excess energy range is almost completely dominated by B1055-52, as we will discuss later. The location of this source close to the Earth, and its low diffusion bubble make its contribution dominant with respect to all the other sources.

The translation of the percentage reported in Fig.~\ref{Fig:top10_percentage} (right panel) into a flux is reported in Fig. \ref{Fig:top10_flux}. Considering the uncertainties, the top 10 overall sources produce a major contribution to the $e^+$ flux independently of the model. 

\begin{table}[t]
\tiny
\begin{center}
\begin{tabular}{ l|l|l|l||l|c|c|c } 
 \textbf{$\mathbf{\tau_0}$ distribution} & \textbf{$\mathbf{\tau_0}$ fixed} & \textbf{Delayed injection}& \textbf{Two zone} & Overall & $d$ [kpc] & $t$ [kyr] & $\dot{E}$ [erg/s]\\
 \hline
\bf{B1055-52} & \bf{B1055-52}  & \bf{B1055-52}  & \bf{B1055-52} & \bf{B1055-52}  & 0.093 & 535.0 & $3 \times 10^{34}$ \\
\bf{J0633+1746} & \bf{J1732-3131} & \bf{J0633+1746} & \bf{J0633+1746} & \bf{J0633+1746}  & 0.19 & 342.0 & $3.2 \times 10^{34}$\\
\bf{B0656+14}  & \bf{J0633+1746}& \bf{B1742-30} & \bf{B1742-30} & \bf{B0656+14} & 0.288 & 111.0 & $3.8 \times 10^{34}$  \\
\bf{B0355+54} & \bf{J2043+2740} & \bf{B0656+14} & B1738-08  & \bf{J1732-3131}  & 0.64 & 111.0 & $1.5 \times 10^{35}$ \\
\bf{J1732-3131}& \bf{B0906-49} & \bf{J2043+2740} & B1749-28& \bf{B0355+54} & 1.0 & 564.0 & $4.5 \times 10^{34}$ \\  
\bf{J2030+4415}& \bf{B0656+14} & B1738-08& J0957-5432& \bf{J2043+2740}  & 1.48 & 1200.0 & $5.6 \times 10^{34}$ \\
B0743-53 &\bf{B0355+54} & \bf{J1732-3131} & \bf{J2030+4415} & \bf{B1742-30} &  0.2 & 546.0 & $8.5 \times 10^{33}$ \\
J1020-5921 & \bf{J0538+2817} & J0954-5430 &B0743-53 & \bf{J2030+4415}  & 0.72 & 555.0 & $2.2 \times 10^{34}$  \\
J0954-5430 & \bf{J2030+4415} & \bf{J2030+4415}&J0945-4833 & \bf{B0906-49} &  1.0 & 112.0 & $4.9 \times 10^{35}$ \\
\bf{B1742-30}& J1016-5819& \bf{B0355+54} &B1001-47 & \bf{J0538+2817} & 1.3 & 618.0 & $4.9 \times 10^{34}$  \\
\end{tabular}
\caption{Left block: list of the top 10 pulsars names that satisfy more frequently the $10^{-4}$ criterion, for each simulation setup. Right block: the top 10 pulsars determined by combining the frequency of occurrence across all models and averaging by the number of realizations that fit the data for each model, together with their distance, age and spin-down energy. The bold text style refers to the top 10 pulsars in column 5.}
\label{tab:table_pulsars}
\end{center}
\end{table}

\begin{figure*}[t]
    \includegraphics[width=0.49\textwidth]{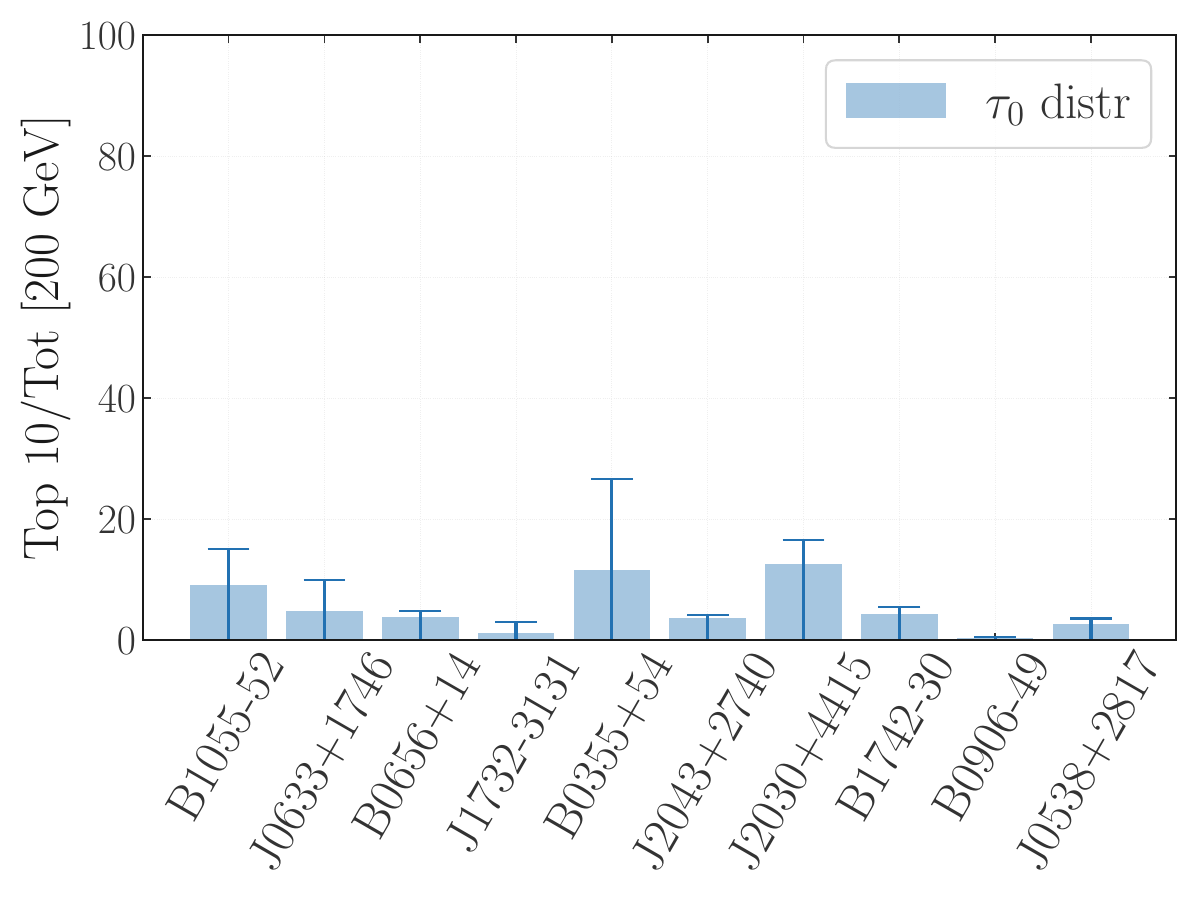}
    \includegraphics[width=0.49\textwidth]{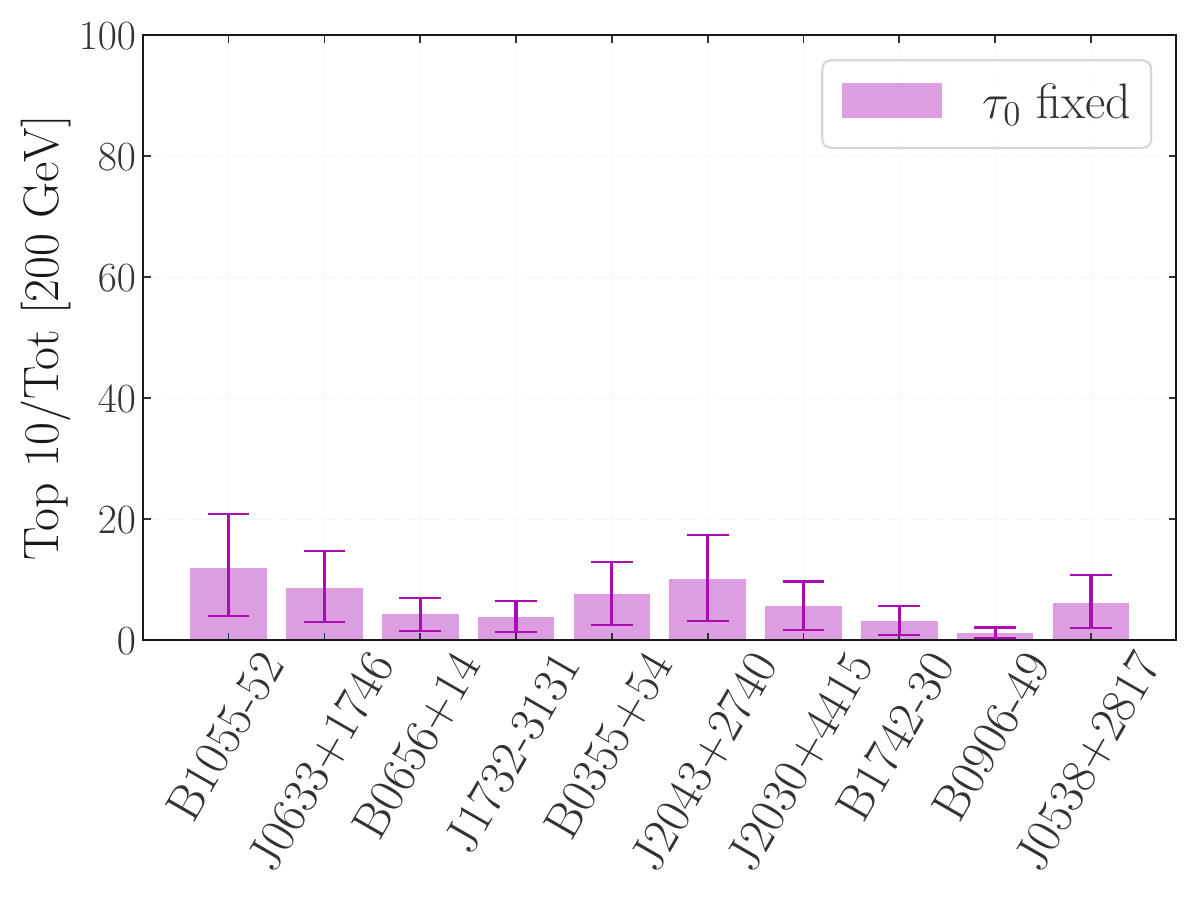}
    \includegraphics[width=0.49\textwidth]{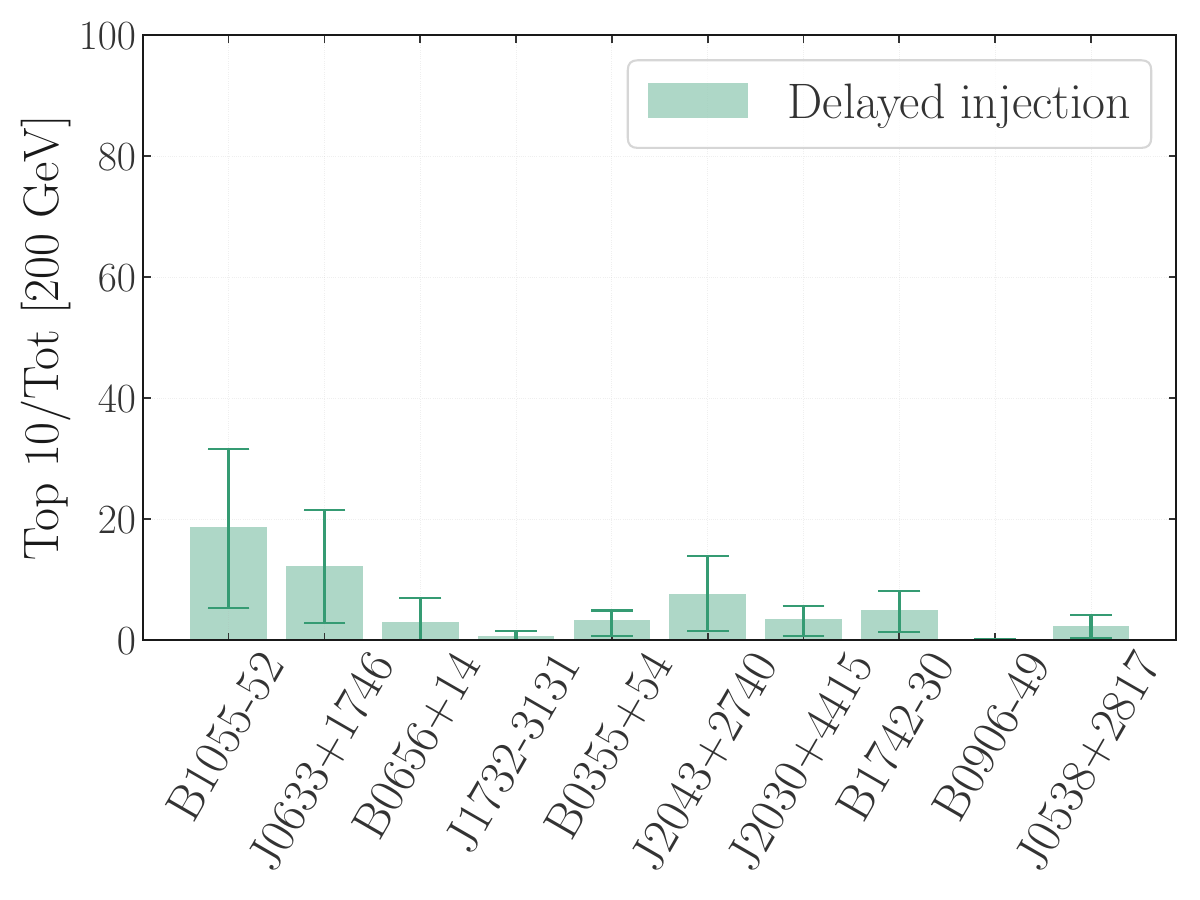}
    \includegraphics[width=0.49\textwidth]{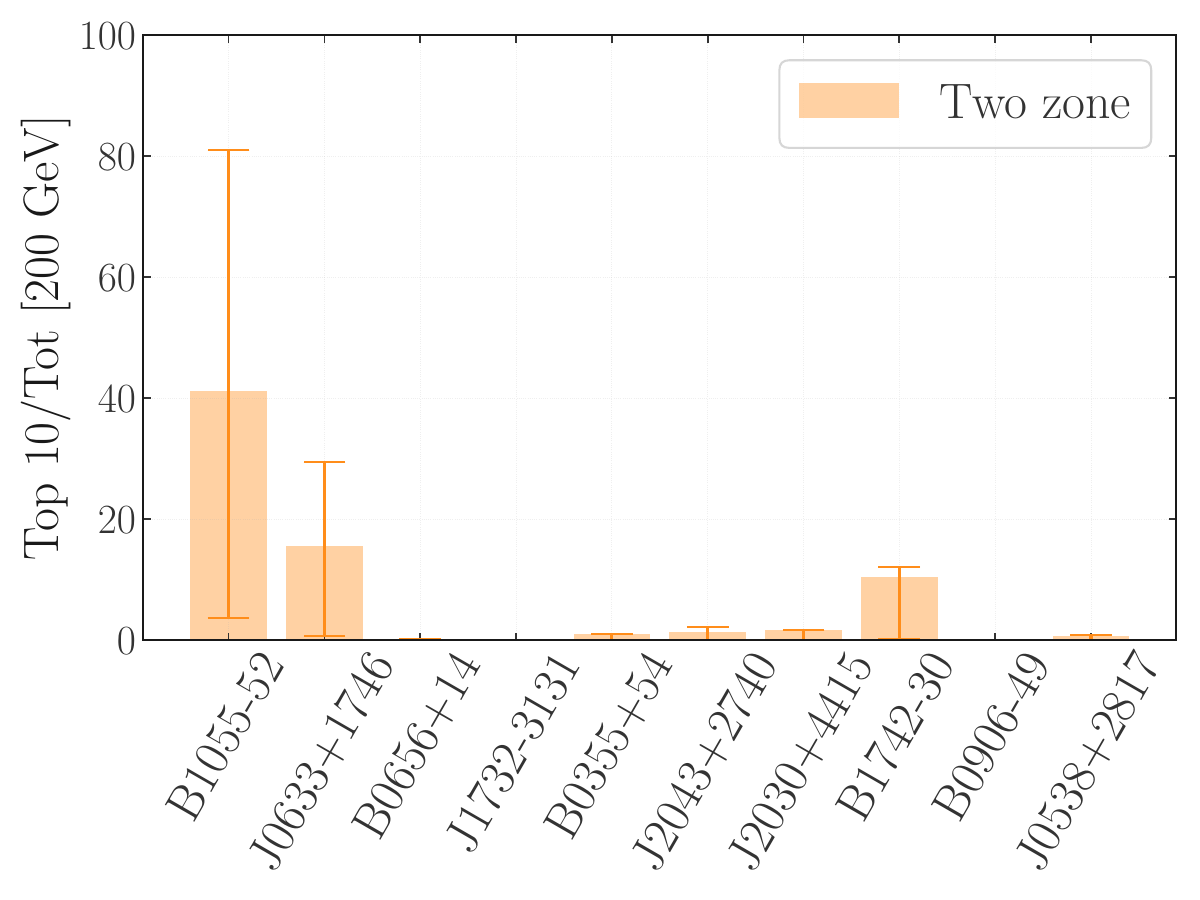}
    \caption{Here are reported the average percentage contributions, together with the 68\% containment bands, from the top 10 brightest sources among the different models at 200 GeV listed in column 5 of Table \ref{tab:table_pulsars}.} 
    \label{Fig:top10_istogram}
\end{figure*}

In Fig.~\ref{Fig:top10_istogram} the average percentage contribution from the top 10 brightest sources among the different models at 200 GeV is shown along  with the 68\% containment band. A clear hierarchy emerges, with B1055-52 and J0633+1746 (Geminga) playing a significant role across all models. B0656+14 (Monogem), although frequently satisfying the $10^{-4}$ criterion, contributes more at higher energies due to its younger age, and is less dominant at 200 GeV.

From Table \ref{tab:table_pulsars}, it is clear how the $e^+$ brightest sources are largely consistent across the different models, except for \textbf{Two zone}, which is mainly dominated by B1055-52. From the characteristics listed in Table \ref{tab:table_pulsars}, it becomes clear that the dominant sources share some key factors: a distance of less than 1 kpc, an age under 600 kyr, and a spin-down power of more than $10^{34}$ erg/s. The leading sources meet all three criteria, highlighting how the dominance is primarily determined by age, distance, and energy output.

\begin{figure*}[t]
    \includegraphics[width=0.49\textwidth]{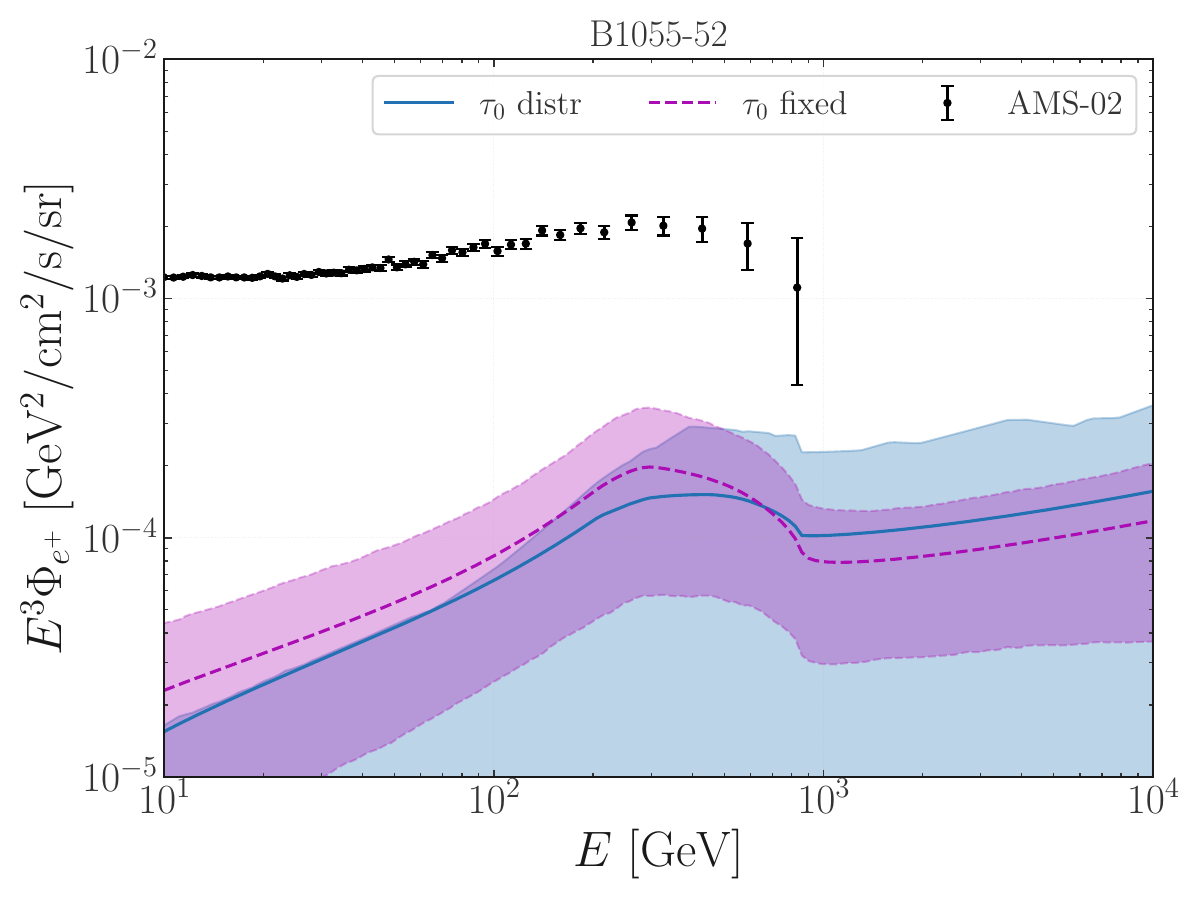}
    \includegraphics[width=0.49\textwidth]{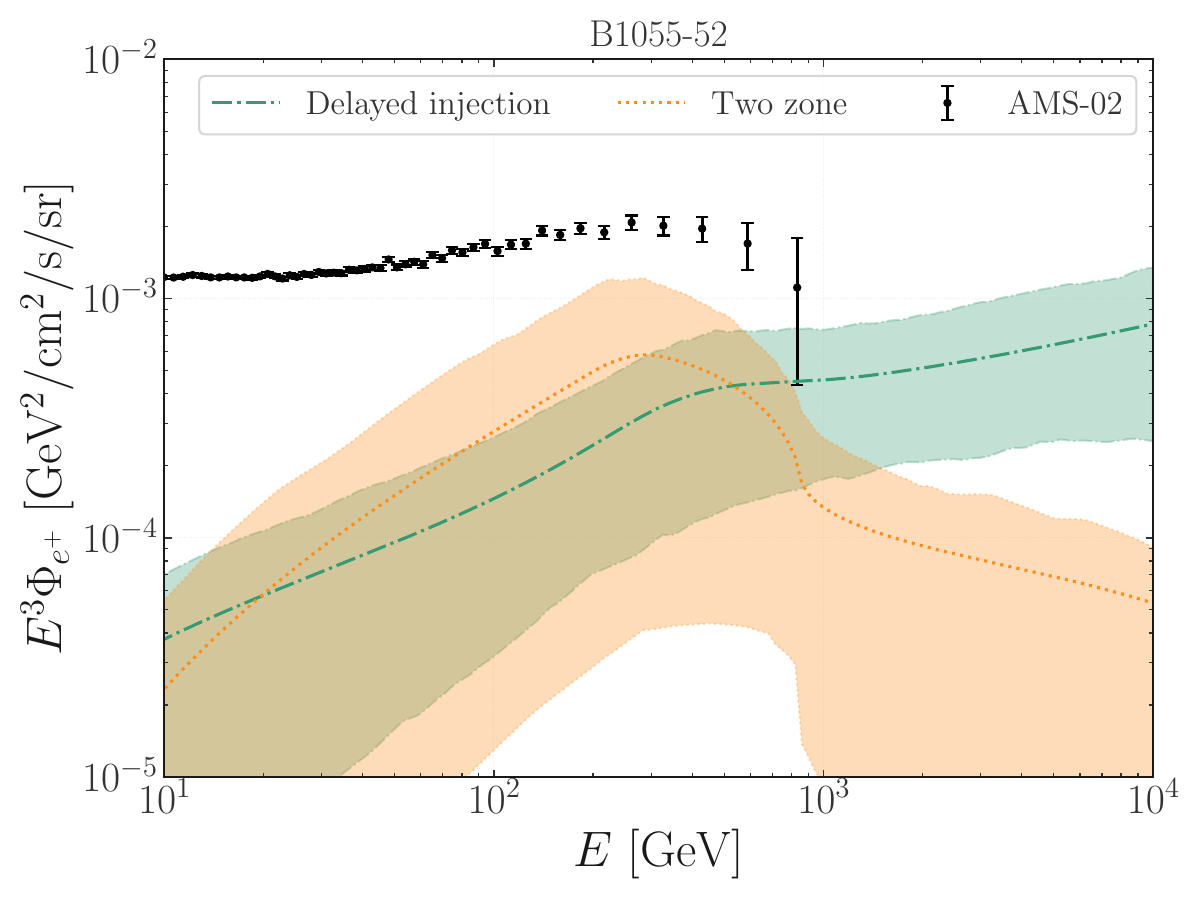}\\
    \includegraphics[width=0.49\textwidth]{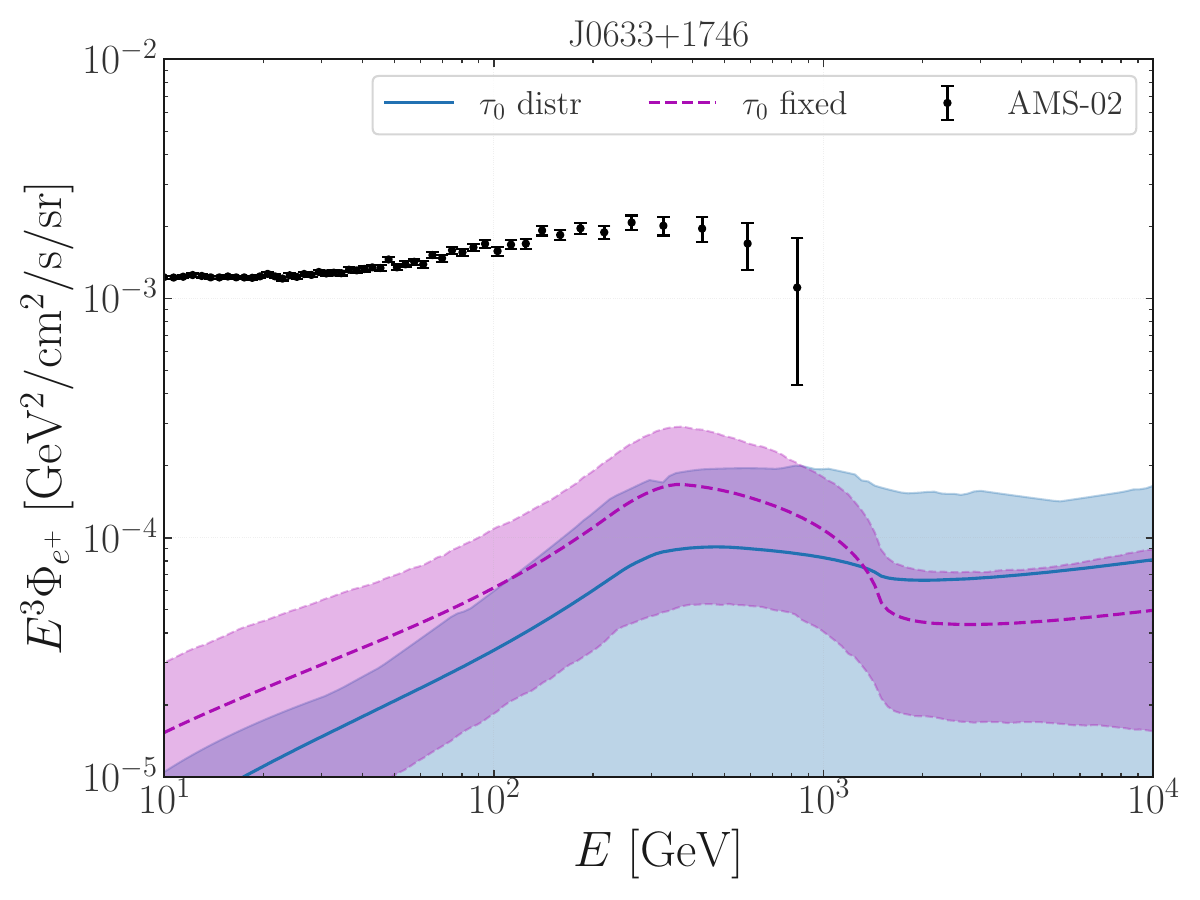}
    \includegraphics[width=0.49\textwidth]{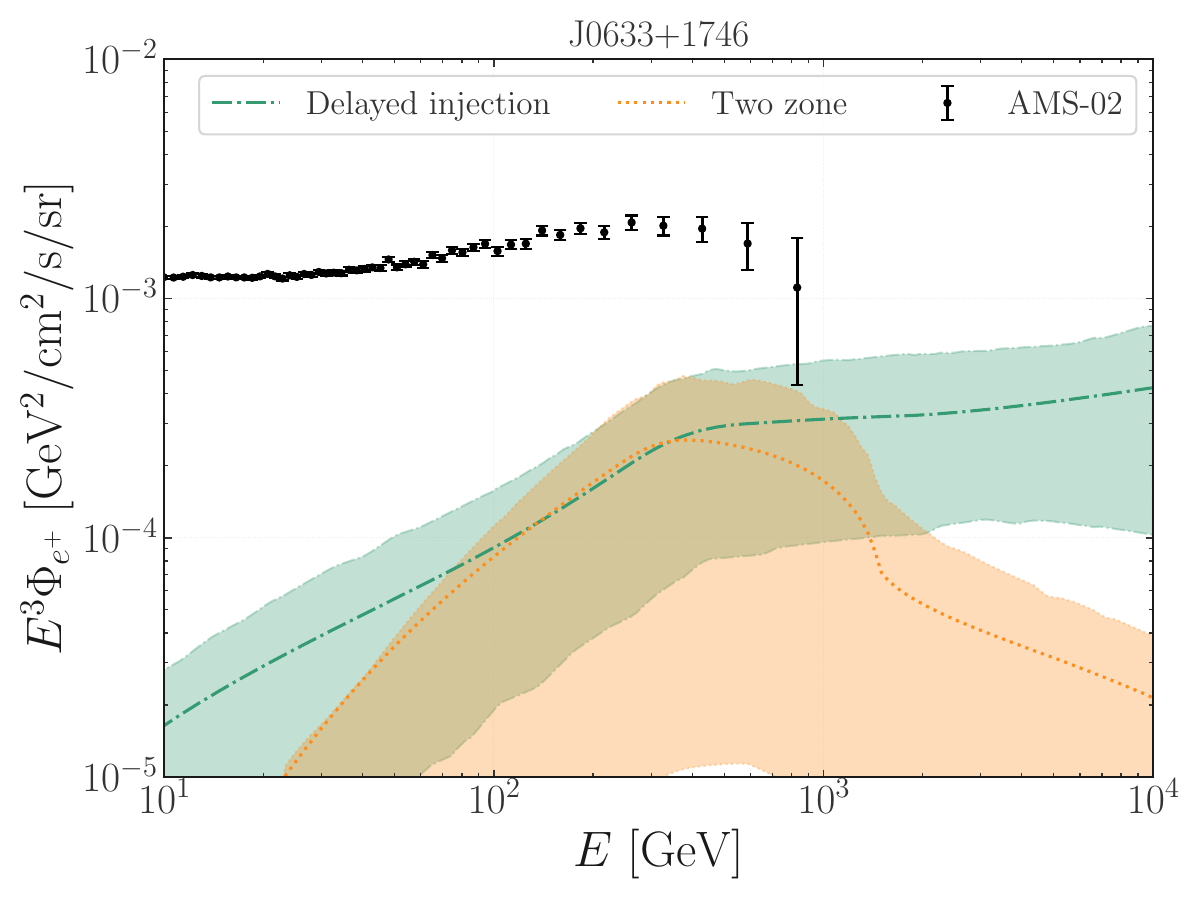}\\
    \includegraphics[width=0.49\textwidth]{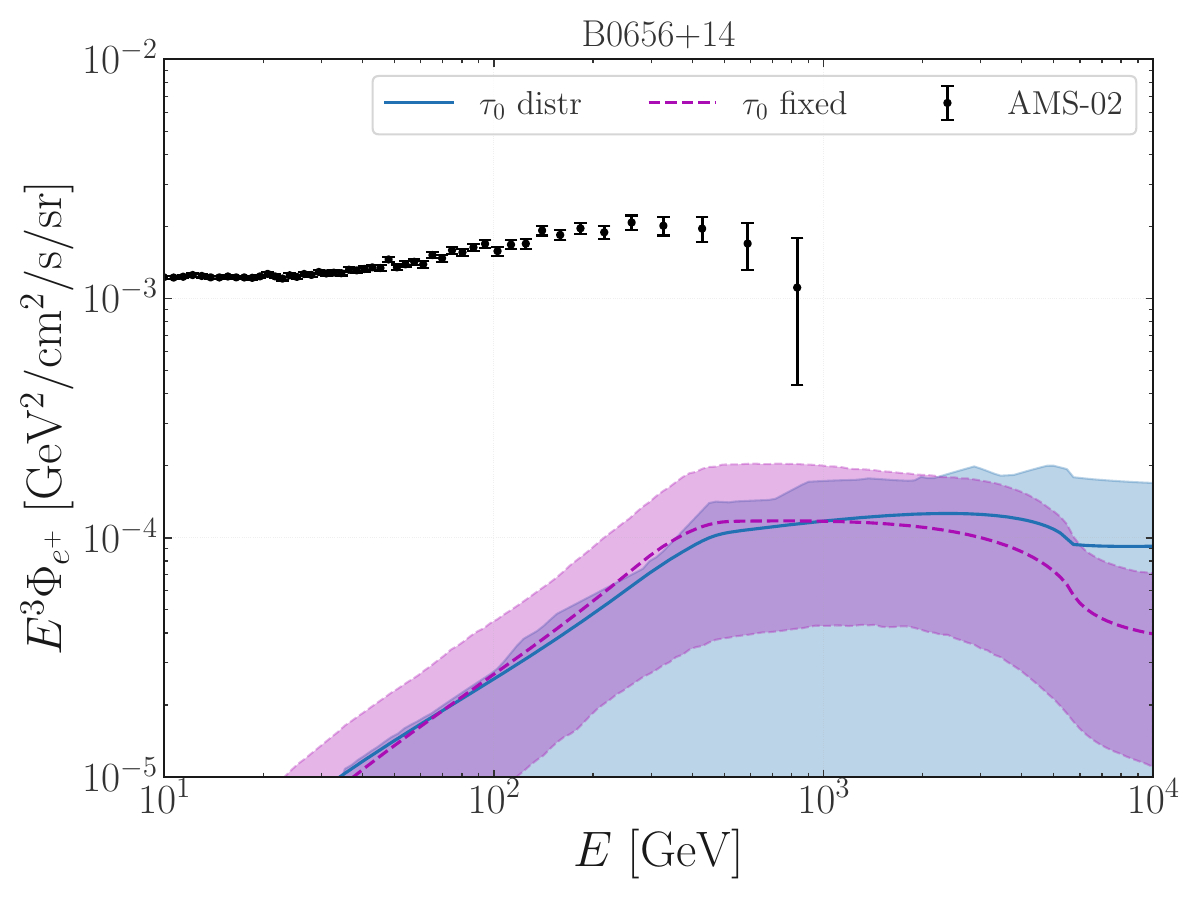}
    \includegraphics[width=0.49\textwidth]{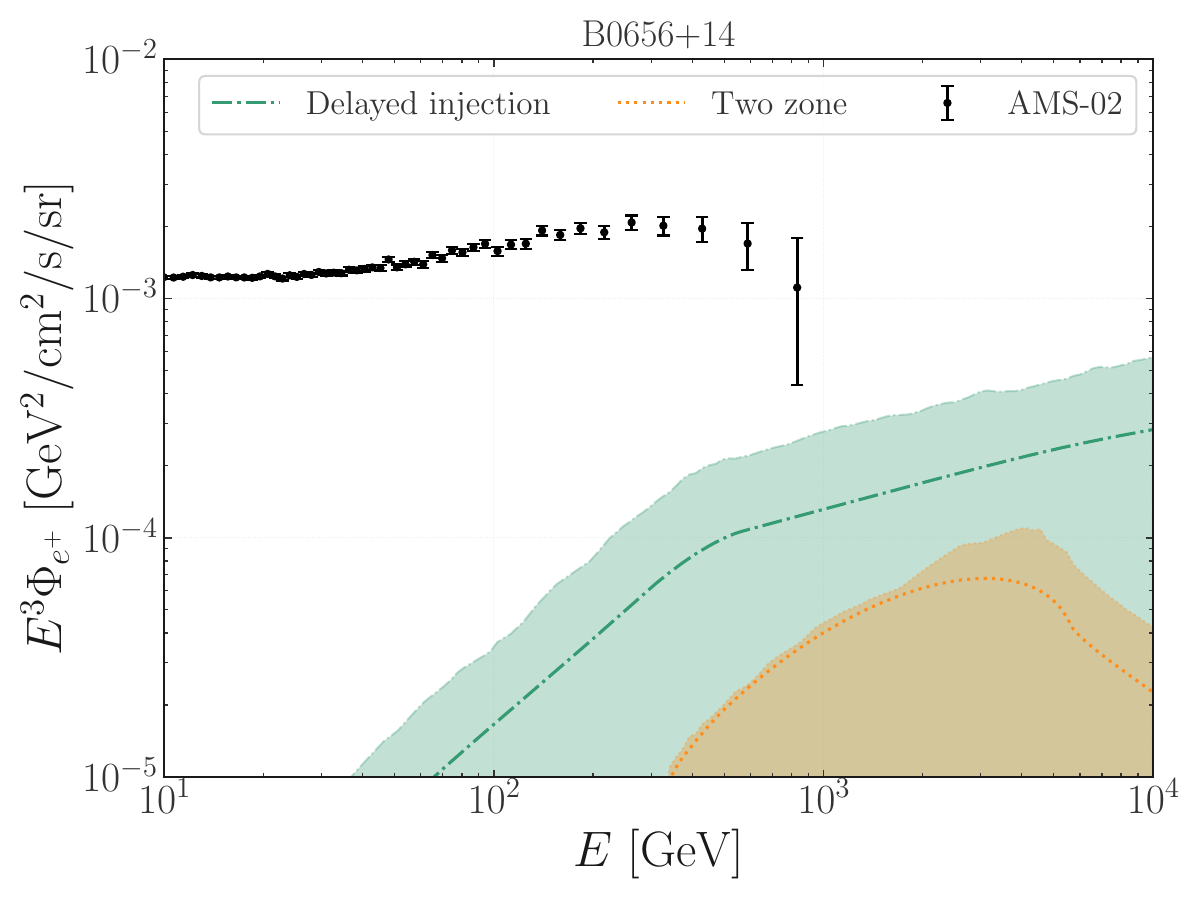}
    \caption{Average fluxes obtained from the realizations that fit well the data, with the 68\% containment band, for the three sources of the top 10 overall ranking, B1055-52, J0633+1746 (Geminga) and B0656+14 (Monogem), for all the tested model.} 
    \label{Fig:top3_flux}
\end{figure*}

In Fig.~\ref{Fig:top3_flux} the average fluxes obtained from the realizations that fit well the data are reported, with the 68\% containment band, for the three sources of the top 10 overall ranking, B1055-52, J0633+1746 (Geminga ) and B0656+14 (Monogem), for all the tested models. For Geminga it is possible to directly compare the predicted flux obtained with the \textbf{Two zone} model with the results shown in \cite{DiMauro:2019yvh}, where the flux from this source was constrained by the results found with a multi-wavelength analysis performed on the Geminga pulsar halo. The flux in \cite{DiMauro:2019yvh} turned out to be higher with respect to our prediction, but it was obtained with different spectral index of the injection spectrum of the pulsar, and a different size of the bubble. Those results are still compatible with our new prediction within the uncertainty band.
Fig.~\ref{Fig:top3_flux} shows how the flux coming from these ''three musketeers'' produces a high contribution in all the four setups, with B1055-52 producing an almost dominant flux within the \textbf{Two zone} model.

\subsection{The anisotropy of the $e^+$ flux }

\begin{figure*}[t]
 \begin{center}
    \includegraphics[width=0.6\textwidth]{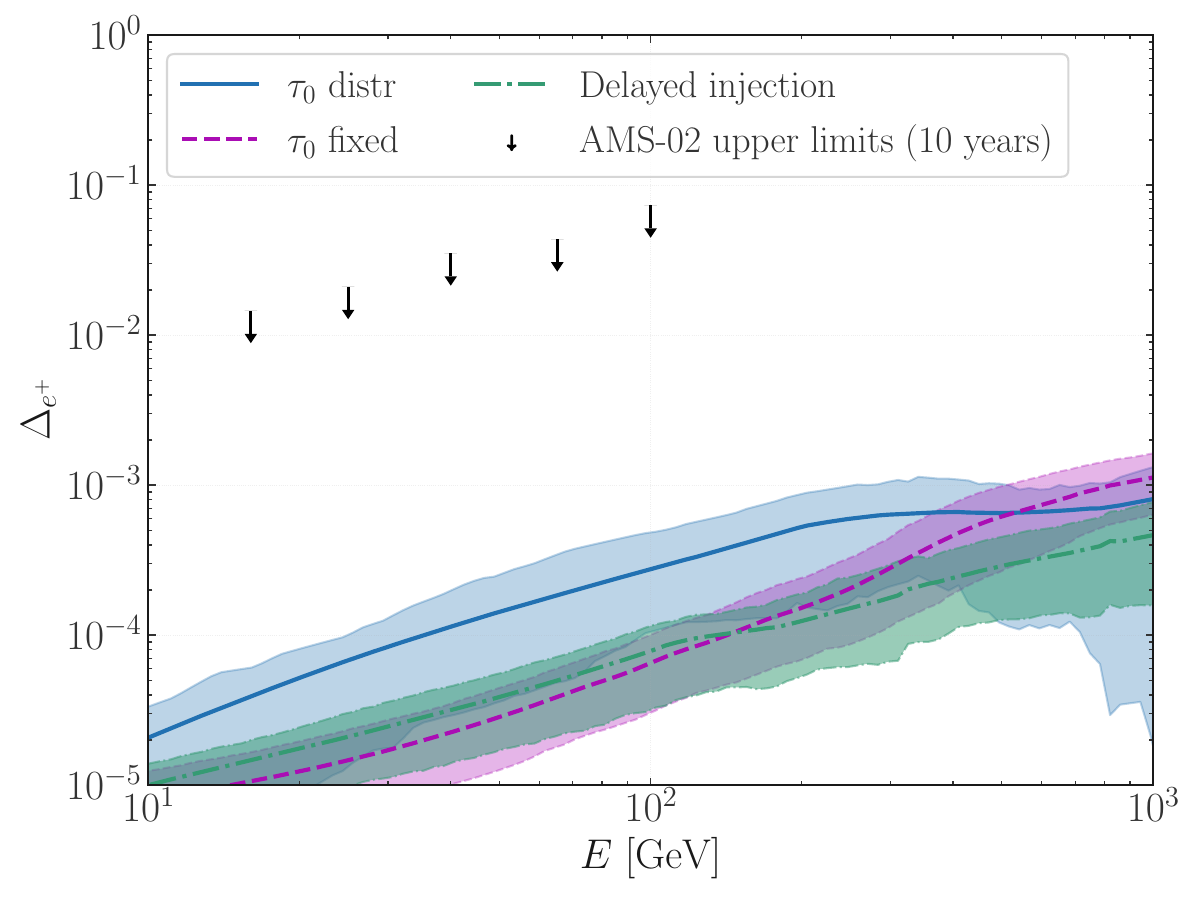}
    \caption{Dipole anisotropy for the realizations that produce a good fit to the data for \textbf{$\mathbf{\tau_0}$ distribution}, \textbf{$\mathbf{\tau_0}$ fixed} and \textbf{Delayed injection}. We compute the anisotropy for each realization, and we provide the average value together with the 68\% containment band. We also show the AMS-02 upper limits of  $\Delta_{e^+}$ obtained from the 10 years data.} 
    \label{Fig:dipole_anistropy}
 \end{center}
\end{figure*}

The anisotropy of the $e^+$ flux could in principle be a valid complementary 
observable in order to understand some physical properties of the emitting sources \cite{Manconi:2016byt}.
We show in Fig. \ref{Fig:dipole_anistropy} the results of the total $e^+$ dipole anisotropy resulting from the contribution of all sources for the realizations that produce a good fit to the data for \textbf{$\mathbf{\tau_0}$ distribution}, \textbf{$\mathbf{\tau_0}$ fixed} and \textbf{Delayed injection} models. We compute the anisotropy for each realization, and provide the average value together with the 68\% containment band.
We predict an anisotropy level between $10^{-4}$ and $10^{-3}$ in the whole energy range. \textbf{$\mathbf{\tau_0}$ distribution}, which is the model that more frequently present the domination of a few pulsars, produce a higher anistropy level with respect to the others, but still well below the AMS-02 upper limits of  $\Delta_{e^+}$ obtained from the 10 years data \cite{VELASCO2024}, 
which stand at least
two orders of magnitude above our predictions. 
It is therefore evident that the properties of ATNF pulsars as emerging from AMS-02 flux data are very unlikely be tested by present or forthcoming data on the $e^+$ anisotropy. 
Our results are similar to the ones found in \cite{Manconi:2016byt}, but with an updated comparison to available experimental upper limits, and using model tuned to the recent insights on $e^\pm$ emission coming from multi-wavelength observations.

\section{Status and perspectives of multi-wavelength insights for the most relevant sources}
\label{sec:multi-w}
These top 10 sources listed in column 5 of Table \ref{tab:table_pulsars} are identified as candidates that could significantly contribute to the $e^+$ flux at Earth across all setups in our investigations of catalog pulsars. We provide here a summary of their main multi-wavelength observations, focusing on those that could help more to constrain the properties of the injected $e^\pm$.

\subsection{The "three musketeers"}
The first three spots of the list provided in column 5 of Table \ref{tab:table_pulsars} are occupied by the three middle-aged pulsars B1055–52, B0656+14, and J0633+1746, that share similar spin-down properties and have been dubbed the 'three musketeers' due to their bright, soft X-ray and high-energy $\gamma$-ray emissions \cite{Becker:1999kk}.

According to the latest version of the ATNF catalog, the pulsar B1055-52 (J1057-5226) is located at a distance of 0.093 kpc. However, the true distance to B1055-52 remains uncertain, varying between 714 and 93 pc depending on the Galactic free-electron density models used: \cite{Cordes:2002wz} and \cite{Yao_2017}, respectively, despite both using the same dispersion measure. This pulsar is highly energetic, with a spin-down power of $3 \times 10^{34}$ erg/s and a spin-down age of 535 kyr. Even if a distance of 714 pc is considered, which would reduce the expected $e^+$ flux, the combination of its $\dot{E}$ and $t$ values suggests that B1055-52 would likely still rank among the top 10 most significant sources when compared to the characteristics of the other leading pulsars reported in Table \ref{tab:table_pulsars}. Alongside Geminga and Monogem, B1055-52 is one of the most studied nearby middle-aged pulsars, showing $\gamma$-ray, X-ray, and radio emissions, as highlighted in a recent multi-wavelength study \cite{Posselt_2023}.
The pulsar is associated with the point-like $\gamma$-ray source 4FGL J1057.9-5227 in the Fermi-LAT 4FGL catalog \cite{Abdollahi_2020}, showing emissions up to a few GeV. No TeV emissions  have been detected thus far. A significant X-ray flux has been observed, including in the recent eRASS:4 catalog obtained by SRG/eROSITA in the 0.2-2.3 keV energy band \cite{Mayer:2024eum}. Although prominent PWN emission has not been detected in the X-ray band, faint extended emission in a small annulus around the pulsar, possibly linked to its PWN, has been observed by Chandra \cite{Posselt:2015bra}.
The possible detection of a $\gamma$-ray halo around this pulsar is very challenging with GeV data from Fermi-LAT because, due to the proximity of the source, the halo could span several tens of degrees or even cover the entire sky. Detecting a signal in X-ray data is also likely prohibitive due to the typically small field of view of detectors. However, at multi-TeV energies, this source can be detected as an extended source a few degrees in size. It is not within the field of view of HAWC or LHAASO, but it can be detected by HESS.

Geminga is a middle-aged pulsar located at a distance of 0.19 kpc, with an age of 342 kyr and a spin-down power of $3.2 \times 10^{34}$ erg/s. A $\gamma$-ray halo has been detected around Geminga at multi-TeV energies by MILAGRO \citep{2009ApJ...700L.127A}, HAWC \citep{Abeysekara:2017science}, and HESS \citep{HESS:2023sbf}, as well as at tens of GeV with Fermi-LAT data \citep{DiMauro:2019yvh}. Despite efforts to detect a synchrotron halo in the X-ray band using XMM, Chandra, and Nustar \citep{Liu:2019sfl,Manconi:2024wlq}, and SRG/\textit{eRosita} \citep{Khokhriakova:2023rqh}, no such emission has been found, with an estimated magnetic field strength of the local ISM of sub-$\mu$G. However, these attempts have limitations, such as narrow energy and spatial coverage \citep{Liu:2019sfl} and short exposure times (400 s) \citep{Khokhriakova:2023rqh}.
Previous studies suggest that Geminga could contribute significantly to the $e^+$ flux, potentially accounting for a few tens of percent of the $e^+$ flux \citep{2009_Hooper,Hooper:2017gtd,DiMauro:2019yvh}, which is consistent with the contribution estimated in this work.
One of the main theoretical uncertainties is related to the size of the halo where the inhibited diffusion takes place. To narrow down these uncertainties, a future study of Fermi-LAT data is needed, where the halo size is estimated alongside the pulsar efficiency and diffusion coefficient.

Monogem is another well-studied, nearby middle-aged pulsar, located at a distance of 0.288 kpc with a spin-down age of 111 kyr and a spin-down power of $3.8 \times 10^{34}$ erg/s. Discovered as a radio pulsar, it exhibits thermal and non-thermal emission across multiple wavelengths, from radio to $\gamma$ rays \citep{Durant_2011,Schwope:2021quk,Fermi-LAT:2022byn}. The properties of the pulsar are consistent with those of the Monogem Ring supernova remnant, an expansive X-ray structure spanning about 25 degrees. An X-ray PWN was identified around the pulsar, with a radius between 0.005 and 0.2 pc, using Chandra data \citep{Birzan_2016}.
This pulsar is younger than Geminga, so it produces, on average, more energetic $e^\pm$. As a consequence, its contribution to the $e^+$ flux is expected to be at the higher end of the observed data. Moreover, the detection of the $\gamma$-ray halo in Fermi-LAT data is challenging due to its angular proximity to Geminga, which is much brighter at GeV energies.

Alongside Geminga, Monogem was found by HAWC to exhibit a  $\gamma$-ray halo extending a few degrees at multi-TeV energies \citep{Abeysekara:2017science}, although only upper limits were obtained when searching for an extended GeV counterpart with Fermi-LAT \citep{DiMauro:2019yvh}. Recent multimessenger studies \citep{Li:2024xwx,DiMauro:2019yvh,Martin:2022hrx} suggest that Monogem contributes only a few percent to the $e^+$ flux.

The potential significance of the "three musketeers" in explaining the $e^+$ excess was recently reiterated in \cite{Fang_2019}. However, their study was limited to a two-zone diffusion scenario and did not account for statistical variation of the injection parameters, nor did it include a fit to the data.

In conclusion, these three pulsars could dominate the contribution to the $e^+$ flux. To verify this hypothesis, a search for their $\gamma$-ray halos is required to estimate the efficiency $\eta$, diffusion coefficient $D_0$, and the size of the inhibited diffusion region. To achieve this, large field-of-view MeV and GeV observations with Fermi-LAT, as well as with future $\gamma$-ray telescopes should be utilized for the analysis of the Geminga halo. In contrast, a dedicated observational campaign by HAWC and LHAASO is needed for Monogem, while HESS data could give insights on B1055-52.

\subsection{Other relevant sources}
In this subsection, we continue discussing the multi-wavelength observations of additional relevant sources identified in Section~\ref{sec:results}. 

The pulsar J1732-3131 is a middle-aged pulsar located 0.64 kpc away, with an age of 111~kyr and a current spin-down power of $1.5 \times 10^{35}$ erg/s. Initially discovered as a radio-quiet $\gamma$-ray pulsar, recent intriguing detections at decameter wavelengths have also been reported \cite{Maan:2017vnh}. This pulsar is positioned near the Galactic center (l=356.307, b=1.007). There is no firm detection of this source at TeV energies, and it is not listed as an extended source in Fermi-LAT catalogs. Additionally, it falls outside the field of view of HAWC.

The pulsar B0355+54, also known as PSR J0358+5413, is a middle-aged pulsar located at a distance of 1 kpc, with an age of 564 kyr and a spin-down power of $4.5 \times 10^{34}$ erg/s. In the TeVcat catalog, there is an entry for a nearby 3HWC source, but this is associated with another pulsar, PSR J0359+5414, which has a significantly larger spin-down power and is slightly younger\footnote{\url{http://tevcat.uchicago.edu/?mode=1&showsrc=420}}. Additionally, in the first LHAASO catalog \cite{LHAASO:2023rpg}, the source 1LHAASO J0359+5406 is located very close to B0355+54, but its association remains unclear, as it may also be related to PSR J0359+5414.

J2043+2740 is a very old pulsar, with an age of 1200 kyr, that has been well-characterized in $\gamma$ rays \cite{Noutsos_2011}. Pulsations from this source have also been detected in radio. Despite being one of the oldest non-recycled pulsars observed in $\gamma$ rays, it remains notably bright with a spin-down power of $5.6 \times 10^{34}$ erg/s, making it a valuable source for constraining magnetosphere models. The ATNF catalog lists its distance as 1.48 kpc. There are no nearby objects associated with it in TeV catalogs, nor is there evidence of nebular emission in X-rays. However, J2043+2740 is X-ray bright and is among the few old, non-recycled, rotation-powered pulsars detected in X-rays \cite{Becker:2004gk}.

The pulsar B1742-30, also known as J1745−3040 is a close (0.2~kpc) middle-aged (546~kyr) pulsar with spin-down power of $8.5 \cdot 10^{33}$ erg/s. This source has been studied together with Geminga, Monogem and other close sources as possible significant contributor to the observed $e^+$ flux, see e.g. \cite{Boudaud:2014dta}. 
Given its sky location close to the Galactic center, the identification of  multi-wavelength counterparts is challenging. 
The pulsar is spatially coincident with a TeV extended source exhibiting a complicated morphology,  HESS J1745–303, located  close to SNR G359.1−0.5, and thus possibly identified as Supernova Remnant/molecular cloud emission, though a leptonic contribution from $e^\pm$ coming from a high energetic pulsar cannot be excluded \cite{HESS:2018pbp}. 

The source J2030+4415 is a 555 kyr $\gamma$-ray pulsar located at 0.72 kpc, with a spin-down power of $2.2 \times 10^{34}$ erg/s. It is very bright in GeV $\gamma$ rays, although no TeV counterpart has been identified so far. Notably, a striking X-ray filament has been observed, which is interpreted as a shock structure where $e^\pm$ are being accelerated to TeV energies \cite{de_Vries_2022,deVries:2020bet}.

B0906-49 is a pulsar located at a distance of 1 kpc, with an age of 112 kyrs and a current spin-down power of $4.9 \times 10^{35}$ erg/s. While no TeV counterpart has been identified in TeV catalogs, it is listed as a point-like source in the Fermi-LAT catalog. A weak extended radio emission, possibly associated with its PWN, has also been detected \cite{Gaensler:1998um}.

The pulsar J0538+2817 was discovered as radio pulsar, and it is usually associated to the supernova remnant SNR S147. It is located at 1.3~kpc, with a characteristic age of 618~kyr and $\dot{E}=4.9 \cdot 10^34$ erg/s.
A faint nebula and pulsation are detected  in X-rays~\cite{Romani:2003ir}, and gamma ray, radio, and H $\alpha$ emission regions are spatially correlated with the remnant~\cite{Tang:2017vxm}. 
Recently, the discovery of a one-sided radio filament within the SNR has been claimed~\cite{Khabibullin:2023eam}, which can be produced by the  escape of energetic $e^\pm$ from the pulsar into the surrounding supernova remnant, constituting a possible radio analogue of the filaments observed in X-rays for the Guitar and Lighthouse nebulae.

\section{Conclusions}
\label{sec:conclusions}

In this paper, we use the high-precision AMS-02 $e^+$ data to constrain key properties of the Galactic pulsar population required to explain the observed CR $e^+$ flux. By modeling the emission from pulsars listed in the latest version of the ATNF catalog \cite{Manchester:2004bp} and fitting their contributions to AMS-02 data \cite{PhysRevLett.122.041102}, we find that a combination of their contribution with the secondary production of $e^+$ emission can account for the majority of the observed $e^+$ flux, especially at high energies.

To predict a pulsar's contribution to the $e^+$ flux at Earth, we use the pulsar's characteristics—such as distance, age, and spin-down power—obtained from the ATNF catalog. Particle emission properties, which are not directly measurable for each pulsar, are modeled using state-of-the-art phenomenological models based on current theoretical knowledge and constrained by multi-wavelength observations of pulsars and their surrounding nebulae and halos. In this analysis, we consider different assumptions for the injection spectrum of $e^{\pm}$, spin-down and particle transport models.

For each realization, we compute the $e^+$ flux at Earth as the sum of a primary component due to pulsar emission and a secondary component from the fragmentation of CRs interacting with the ISM. The results are then fitted to AMS-02 $e^+$ data. We find that a fraction of the  realizations provide an excellent fit to AMS-02 data.

The pulsar contribution models we explored consistently show that the $e^+$ flux is dominated by a small number of nearby, middle-aged pulsars, particularly those within 1 kpc of Earth. Notably, the top 10 brightest pulsars can contribute up to 80\% of the total $e^+$ flux at energies above 100 GeV, reaffirming previous findings that a small number of sources can account for the majority of the $e^+$ excess \cite{Orusa_2021}. 

The implications of these findings are twofold: we reinforce the importance of local, high-energy pulsars in explaining the $e^+$ excess, and we emphasize the need for multi-wavelength follow-up observations of the most important sources listed in this analysis. More precise $\gamma$-ray observations of the extended halos around these pulsars, as well as studies of synchrotron emission in the X-ray band, 
are essential to constrain the injection spectra and $e^+$ diffusion properties, see e.g. the case of Geminga \cite{Manconi:2024wlq}. We provide a list of the 10 most important sources identified across the different tested models, which should be the primary focus of future multi-wavelength analyses.

Specifically, B1055-52, Geminga (J0633+1746), and Monogem (B0656+14) consistently emerge as the primary contributors across all tested models and stand out as prime targets for future $\gamma$-ray observations, particularly with instruments like Fermi-LAT and observatories like HAWC, LHAASO, and HESS, as well as with future TeV observatories like CTA \cite{CTAConsortium:2013ofs}, and SWGO \cite{abreu2019southernwidefieldgammarayobservatory}. 
Regarding X-ray observations, detecting the synchrotron counterpart of the Inverse Compton pulsar halos with current X-ray telescopes is challenging due to the proximity of these sources ($<1$~kpc) \cite{Manconi:2024wlq}. However, future wide field-of-view X-ray telescopes, as well as observations targeting more distant Galactic pulsars, could still provide valuable insights into the properties of the emitted $e^\pm$ on a population-wide scale.

These results provide a solid foundation for future observational campaigns focused on these pulsars, which will be key to refining our understanding of cosmic ray $e^+$ transport and the physical processes governing pulsar emission.

\acknowledgments
We thank H. Hakobyan and S. Recchia for inspiring discussion. LO acknowledges the support of the Multimessenger Plasma Physics Center (MPPC), NSF grants PHY2206607.
SM acknowledges the European Union's Horizon Europe research and innovation program for support under the Marie Sklodowska-Curie Action HE MSCA PF–2021,  grant agreement No.10106280, project \textit{VerSi}.
The work of FD is supported by the 
Research grant {\sl The Dark Universe: A Synergic Multimessenger Approach}, No.
2017X7X85K funded by the {\sc Miur}. FD and MDM~acknowledge support from the research grant {\sl TAsP (Theoretical Astroparticle Physics)} funded by Istituto Nazionale di Fisica Nucleare (INFN).


\bibliographystyle{JHEP}
\bibliography{main}

\end{document}